\documentclass[11pt]{article}

\usepackage[utf8]{inputenc}
\usepackage[T1]{fontenc}
\usepackage{lmodern}
\usepackage[margin=1in]{geometry}
\usepackage{amsmath, amssymb, amsfonts}
\usepackage{graphicx}
\usepackage{subcaption}
\usepackage{booktabs}
\usepackage{siunitx}
\usepackage{chemformula}
\usepackage{microtype}
\usepackage{xcolor}
\usepackage[colorlinks=true,
            linkcolor=blue!60!black,
            citecolor=blue!60!black,
            urlcolor=blue!60!black]{hyperref}
\usepackage[capitalise,noabbrev]{cleveref}

\usepackage[backend=biber,
            style=chem-acs,
            articletitle=true,
            doi=true,
            biblabel=brackets,
            sorting=none,
            maxbibnames=99]{biblatex}
\title{Fine-Tuning a Universal Machine-Learned Interatomic Potential for Oxygen Plasma Interactions with WS\textsubscript{2}}

\author{
  Jaehong Kwon, Andrew S.\ Rosen, and David B.\ Graves\textsuperscript{*} \\[0.5em]
  \normalsize\itshape Department of Chemical and Biological Engineering,
  Princeton University, Princeton, NJ 08544 USA \\[0.3em]
  \normalsize\upshape E-mail: \href{mailto:dgraves@princeton.edu}{dgraves@princeton.edu}
}

\date{\today}

\begin{document}
\maketitle
\thispagestyle{empty} 
\newpage

\thispagestyle{empty} 
\begin{abstract}
\noindent
Molecular dynamics simulation of plasma-surface interactions requires an interatomic potential that is simultaneously accurate, computationally efficient, and able to describe many elements and bonding types in reactive systems. In principle, a foundation model for machine-learned interatomic potential (MLIP) can meet these demands. We explore the use of the Universal Models for Atoms (UMA) model, developed by Meta FAIR, for the interactions of oxygen plasma species on a multilayer of WS$_2$, a promising 2D material. Starting from the pretrained \texttt{uma-s-1p1} model under the Open Catalyst 2020 (OC20) task, we apply an iterative fine-tuning loop.  Even in the absence of fine-tuning, the pretrained model reproduces the production-scale observables of interest, namely, chemisorbed S and O coverage under 15~eV O$^+$ and O$_2^+$ bombardment. These results were obtained without spin polarization and Hubbard $U$ correction. Nonetheless, fine-tuning with spin polarization and a Hubbard $U$ correction reduces the energy and force mean absolute error (MAE) to $4.5\times10^{-3}$~eV/atom and $0.076$~eV/\AA{}, respectively.
\end{abstract}

\newpage
\setcounter{page}{1}  

\section{Introduction}
\label{sec:introduction}
Molecular dynamics (MD) simulation of plasma-surface interactions can reveal atomic scale mechanisms that can be critically important in applications ranging from thin film etching to deposition to surface modification~\cite{graves_molecular_2009}. The growth of the semiconductor and quantum device industries coupled with the move towards features approaching atomic scale motivates the need to control plasma surface interactions at the atomic scale~\cite{oehrlein2024future,graves2023report}. A key challenge in utilizing MD simulation is the development and evaluation of a suitable interatomic potential (IP)~\cite{neyts2017molecular}. The plasma-surface interaction application often poses significant constraints on MD simulations, complicating the development of a suitable IP.

First, chemically reactive plasma will almost always strongly alter its bounding surfaces, and realistic simulations must be capable of reproducing this effect. A common rule of thumb is that for a single set of process conditions a steady-state surface requires about $10^{16}$ ion impacts per cm$^2$~\cite{roh_sputtering_2024, graves_molecular_2009}. Often, neutral radical fluxes are also important, and typical radical-to-ion flux ratios range from tens to several hundreds. Simulations require a finite-size cell, usually consisting of a surface area of about $10~\mathrm{nm}^{2}$ and a depth of about $1$--$5~\mathrm{nm}$ depending on the impacting ion energy~\cite{neyts2017molecular, kounis-melas_deep_2025}. For ions below about $200~\mathrm{eV}$, cell depths of about $5~\mathrm{nm}$ are usually sufficient~\cite{humbird_molecular_2007, kubota_molecular_1998, vella_near-surface_2023}. Simulation cells of this size include on the order of several thousand atoms~\cite{neyts2017molecular, kounis-melas_deep_2025}.

A typical study of a single set of conditions (e.g. ion energy and flux, neutral and ion composition) requires enough impacts to make meaningful comparisons to experiments. Hundreds to tens of thousands of impacts may be necessary for systems of this size, which amounts to a cumulative simulated time ranging from sub-nanosecond to several tens of nanoseconds~\cite{graves_molecular_2009, neyts2017molecular}. However, for a proper understanding of the applications of interest, it is usually necessary to simulate a relatively wide range of process conditions with varying species energy, angle, fluxes, composition, etc. The result of these features of MD simulations of plasma-surface interactions is that the computational cost of a single trajectory cannot be too large, or the simulation becomes impractical.

Another often critically important issue is the ability to simulate the effects of many different species from the plasma, interacting with often complex surface material. One example is dielectric etching of silicon dioxide and silicon nitride with etch gases that typically include carbon, hydrogen, fluorine, among others. The IP must therefore be capable of modeling chemical interactions between at least six different atoms with multiple types of chemical bonds~\cite{vanraes2021multiscale}.

The challenge of developing a relatively computationally efficient IP that can also incorporate many different atoms and bonding types has limited the use of MD in modeling industrially important plasma-surface applications to date.  However, recent dramatic developments in computational material science involving machine-learned IPs (MLIPs) promise to overcome these limitations~\cite{ko_recent_2023, jacobs2025practical}. The present paper focuses on developing an MLIP suitable for modeling oxygen plasma-surface interactions for a multilayer film of the 2D transition metal dichalcogenide WS$_2$.

Transition metal dichalcogenide (TMD) materials, such as WS$_2$, are gaining attention as a potential complementary building block to silicon integrated circuits in semiconductor chips to address scaling problems in conventional silicon devices~\cite{wang_two-dimensional_2022}. These materials typically form a series of stacked 2D layers, held together by relatively weak attractive dispersion forces rather than relatively strong covalent bonds, which allow for atomically sharp geometry, layer tunable band gap, and dangling-bond-free interfaces. This configuration allows for thin materials with desirable electronic and optical properties~\cite{wang_two-dimensional_2022,cao_future_2023,kim_future_2024}. Among the variety of TMD materials, WS$_2$ is of particular interest due to its thermodynamically stable 2H phase~\cite{wilson_transition_1969,jain_commentary_2013,horton_accelerated_2025,sun_thermodynamic_2016} and its single layer band gap of approximately 1.8 to 2.1 eV~\cite{kuc_influence_2011,bui_first-principles_2015}. 

However, because the electronic properties of TMD materials are strongly dependent on the number of layers, the ability to precisely control over layer count is critical for device performance. Atomic layer etching (ALE), where layers are removed one layer at a time through a cyclic process step of chemical modification and selective removal, can be a viable pathway for layer-by-layer control~\cite{kanarik_overview_2015,kanarik_predicting_2017,t_carver_atomic_2015}. For WS$_2$, one possible process is to use an oxygen plasma to first oxidize the top layer, followed by a second plasma step to selectively remove the resulting oxide. In this paper, we focus on the first half-step, using oxygen plasma to oxidize WS$_2$ to form a WO$_x$ layer. 

You et al. demonstrated an oxidation of a multilayer WS$_2$ with a remote O$_2$ plasma and removed the oxide in a KOH solution rather than with a second plasma~\cite{you_editors_2023}. Their vertical etch rate of about 0.65~nm per cycle is close to the thickness of a single WS$_2$ monolayer, indicating that the oxidation was confined to the topmost layer and was therefore self-limiting. Kang et al. treated single-layer MoS$_2$ in an O$_2$/Ar plasma and observed the Mo$^{6+}$ peak of MoO$_3$ domains in XPS~\cite{kang_photoluminescence_2014}. These studies differ from the present work in the species that reach the surface. You et al. used a remote plasma, where the surface sees mainly reactive neutrals. Kang et al. studied MoS$_2$ and did not report an ion energy. Neither demonstrates the response of WS$_2$ to oxygen ions with a known energy, but they still lead us to expect oxidation of WS$_2$ to WO$_x$. The primary focus of this paper is to explore the application of MLIPs for MD simulations of plasma-TMD interactions and we use this process as an example. 

As noted above, MLIPs can achieve near-DFT-level accuracy but with a considerably lower computational burden~\cite{ko_recent_2023, zhang_deep_2018, kounis-melas_deep_2025}. MLIPs can be most commonly classified into local MLIPs (L-MLIPs), which build the per-atom energy from fixed descriptors of the local environment, and graph MLIPs (G-MLIPs), which learn that representation through message passing over a graph of the atomic structure~\cite{ko_recent_2023}. In a G-MLIP, information propagates beyond the neighbor-list cutoff through successive message-passing layers, so the effective receptive field is roughly the number of layers times that cutoff. L-MLIPs include the Behler-Parrinello neural network potential~\cite{behler_generalized_2007}, Gaussian approximation potential~\cite{bartok_gaussian_2010}, and Deep Potential MD or DeePMD~\cite{zhang_deep_2018}. These alternatives have been shown to accurately learn the potential energy surface from DFT training data. More recent G-MLIP architectures, specifically equivariant graph neural networks like NequIP \cite{batzner_e3-equivariant_2022} and MACE \cite{batatia2022mace}, are reported to improve accuracy and data efficiency.

Both types of MLIPs have been used in the MD studies of plasma-surface interactions, with L-MLIPs trained on system-specific DFT data dominating the earlier work. For instance, Kounis-Melas et al.\ demonstrated that DeepPot-SE yields good agreement with the experimental measurements in different etching conditions of Si-Cl$_2$-Ar$^+$~\cite{kounis-melas_deep_2025}, and Hong et al.\ trained a neural-network potential for HF etching of Si$_3$N$_4$~\cite{hong_atomistic_2024}.

Building on G-MLIP architectures, foundation models emerged recently as a notable advance. Foundation models are trained on chemically diverse datasets spanning the periodic table. Representative examples include MACE-MP-0~\cite{batatia_foundation_2025}, MatterSim~\cite{yang_mattersim_2024}, and the Universal Models for Atoms (UMA) family released by Meta FAIR~\cite{wood_uma_2026}. Oh et al.\ applied a knowledge distillation on the SevenNet-Omni foundation model, and its student model was used to simulate plasma etching of SiO$_2$ with CF$_x$ plasma. Using this approach, these authors reproduced experimental etch yields~\cite{oh_lightweight_2026}. However, there are not many studies on applying a foundation MLIP to plasma-surface interactions due to its recent emergence. The public leaderboards that rank pretrained models, such as the Open Catalyst Project leaderboard~\cite{ocp_leaderboard} and Matbench Discovery~\cite{riebesell2025framework}, score them on relaxation and crystal-stability benchmarks, but a high ranking there does not indicate that a model is suitable for plasma-surface simulation.

Simulating the plasma-surface interaction can be more challenging than the near-equilibrium systems for which foundation MLIPs were built. Energetic bombardment continuously drives the top of the film away from the starting material, from pristine WS$_2$ toward an amorphized WO$_x$ layer over the WS$_2$ underneath. Therefore, a trajectory can leave the domain of the training set from the pretraining step. In this work, we fine-tune the UMA-S foundation MLIP (described in Sec.~\ref{sec:methods}) for interactions of oxygen plasma species with WS$_2$ surfaces, assessing the performance of both the pretrained and fine-tuned models. The species considered are O$^+$ and O$_2^+$ ions at normal incidence~\cite{hsu_comparison_2006}. We neglected the effects of O radicals for simplicity. We find that the pretrained UMA-S model captures the plasma-surface observables of interest, namely, the chemisorbed S and O coverage. Iterative fine-tuning improves upon this and further reduces the mean absolute error (MAE). The same workflow therefore can be used to achieve higher accuracy for systems where the pretrained model with no additional training is not sufficient. The remaining sections of the paper are organized as follows: Section~\ref{sec:methods} describes simulation and more detailed fine-tuning methodology; Section~\ref{sec:results} presents the performance of the fine-tuned models; and Section~\ref{sec:conclusion} presents the summary of the main findings and the outlook.

\section{Methods}
\label{sec:methods}
The present work employs UMA, which blends massive training data with efficient inference time, making it a strong fit for the plasma-surface models. Specifically, the UMA model was trained on approximately 500 million atomic configurations across diverse domains including materials, catalysts, molecules, etc., while utilizing a Mixture of Linear Experts (MoLE) strategy that enables large enough model capacity without sacrificing the inference speed~\cite{wood_uma_2026}. Across different benchmarks, the pretrained UMA model has been shown to match or surpass the accuracy of the models trained for specific applications~\cite{wood_uma_2026}. Using UMA, however, requires selecting a task, or a specific chemical domain, each trained with different training sets and DFT settings~\cite{wood_uma_2026}. For a periodic, inorganic system such as W--S--O, the only two candidates among the five tasks available in \texttt{uma-s-1p1} are the Open Catalyst 2020 (OC20) and Open Materials 2024 (OMat) tasks~\cite{chanussot_open_2021,barros2026open}. The other tasks cover non-periodic molecular systems, organic molecular crystals, and metal--organic frameworks, none of which describes a periodic inorganic surface interacting with incoming gas-phase species ~\cite{wood_uma_2026}.

UMA's available tasks are constrained by their training data in two ways. First, neither the OC20 nor the OMat task has seen configurations generated by bombardment with energetic species (around 15 eV), which are characteristic of plasma-surface interactions~\cite{chanussot_open_2021,barros2026open}. OMat additionally lacks the gas-surface interactions central to our system~\cite{barros2026open}. Second, the DFT parameters for both tasks can be inappropriate for the specific chemistry. Specifically, OC20 omits both spin polarization and a Hubbard $U$ correction~\cite{chanussot_open_2021}, yet both matter for W--S--O chemistry. The ground state of O$_2$ is a triplet, and its dissociation dynamics at surfaces change qualitatively depending on its spin state~\cite{behler2005dissociation,behler2008nonadiabatic}. As the surface oxidizes, plain semilocal DFT (like Perdew-Burke-Ernzerhof (PBE)~\cite{perdew1996generalized} or revised Perdew-Burke-Ernzerhof (RPBE)~\cite{hammer1999improved}) no longer captures the W \textit{5d} states of WO$_x$ well, so an on-site Hubbard $U$ correction becomes necessary~\cite{wang2011electronic,bondarenko2015polaron}. The spin polarization also becomes important when the high-energy impact creates under-coordinated W atoms resulting in local magnetic moments~\cite{bondarenko2015polaron,gerosa2016anisotropic}. These considerations suggest that a spin-free, $U$-free model could lose accuracy as oxidation proceeds, which is one of the things this work sets out to test. Fine-tuning addresses this issue by further training the pretrained model on target-chemistry configurations labeled at an appropriate level of theory, here PBE+D3+$U$+spin.

When fine-tuning is needed, it requires generating a new training set and retraining the pretrained model, but it is still expected to be more efficient than training a dedicated potential from scratch. Because the pretrained model encodes broadly transferable representations learned from millions of diverse configurations, fine-tuning starts from a far stronger baseline than random initialization of parameters. It therefore requires fewer labeled configurations to reach a given accuracy~\cite{ko_recent_2023}, and it is likely to generalize more robustly to configurations that lie outside the fine-tuning set but within the scope of the pretraining data. By contrast, a model trained from scratch on a limited set has no such fallback~\cite{liu_fine-tuning_2026}.

The computational workflow in the present study has two parts: MD simulation of O$^+$ and O$_2^+$ bombardment on a model WS$_2$ slab and an iterative fine-tuning loop of MD exploration, configuration sampling, DFT labeling, and fine-tuning on the resulting reference data. All MD simulations in this work use Meta's pretrained UMA foundation model~\cite{wood_uma_2026}, specifically the \texttt{uma-s-1p1} variant. The model uses a 6~\AA{} neighbor-list cutoff and four message-passing layers, giving an effective receptive field of roughly 24~\AA{}. Despite having 150 million total parameters, UMA-S activates only 6 million per inference through its MoLE architecture, giving an inference speed of 16 MD steps per second for 1{,}000-atom systems on a single H100 GPU, which is tractable for our simulations~\cite{wood_uma_2026}. The inference speed of the next larger variant UMA-M was too slow given our compute budget. The OC20 task was chosen over the OMat task after comparing the pretrained UMA-S model's performance to the AIMD reference data on 15~eV O$^+$ impact on the pristine WS$_2$ surface. Both tasks have shortcomings for this system, but OC20's are easier to correct with fine-tuning data. The detailed explanation for the choice to use OC20 can be found in Sec.~\ref{app:oc20-omat}. Portions of the analysis and simulation code were written with the assistance of Claude Code (Anthropic). All AI-assisted code was reviewed, tested, and validated by the authors, who take full responsibility for its correctness and for the results reported in this work.

\subsection{Simulation setup}
\label{sec:methods:sim}
MD simulations or evaluation of energies and forces are performed in the Atomic Simulation Environment (ASE, version 3.28.0)~\cite{hjorth2017atomic}. When the fine-tuned model was used, no dispersion correction was added, since a D3 van der Waals correction~\cite{grimme2010consistent} was added to the training set. When the pretrained model was used, a D3 van der Waals correction was added through the \texttt{torch-dftd} library~\cite{takamoto2021pfp}. The following explains the core assumptions in these plasma-surface modeling and the simulation workflow.
 
\paragraph*{Core assumptions of the plasma-surface model}
Before discussing the simulation workflow, it is worthwhile to first discuss two major assumptions of the plasma-surface interactions. First, we treat the incoming ions as fast neutrals. We make this assumption because although the dominant incoming species from the plasma are O$^+$ and O$_2^+$, charge transfer from the ion to the surface is essentially complete by the time the ion is 2--3~\AA{} above the surface~\cite{hagstrum_theory_1961}. Second, because the incoming ion species have relatively high kinetic energy, they will generally overcome any activation energy barriers. The reaction kinetics is prompt and is completed within the collision time. This is generally less than or on the order of one or two picoseconds (10$^{-12}$ s). Following the impact trajectory simulation of 1 ps, we implement a post-impact stratified Berendsen thermostat to remove excess kinetic energy due to the high-energy impact~\cite{lemak1994berendsen,humbird_computational_nodate}.
 
\paragraph*{Simulation workflow}
We begin with the 2H-WS$_2$ bulk structure~\cite{jain_commentary_2013,horton_accelerated_2025}, generated through Pymatgen~\cite{ong2013python}. Then the bulk cell is relaxed and replicated into a $6\!\times\!6\!\times\!3$ slab with periodic boundary conditions in the lateral directions. The bottom $2.0~\text{\AA}$ of the slab is fixed to stabilize the cell, and the mobile atoms are thermalized with Langevin dynamics at $300~\mathrm{K}$ before the first impact.

Each impact cycle begins by placing an incoming ion above the surface at a random lateral position, with the velocity in the z-direction pointing downward with the magnitude calculated from the chosen kinetic energy. We use an impact energy of 15~eV throughout, chosen to represent the companion experiment~\cite{DonnellyPC}. The electron temperature measured there is about 3~eV, and ions crossing the sheath to a surface at floating potential acquire an energy of approximately five times the electron temperature, or roughly 15~eV. For O$_2$, the molecular orientation is also randomized. We simulate the collision under the microcanonical (NVE) ensemble for 1 ps with a 0.5 fs timestep and then cool the system back to 300 K with a stratified Berendsen thermostat~\cite{lemak1994berendsen,humbird_computational_nodate}. Conventional plasma-surface MD studies typically use a 3D bulk substrate and layer cooling is usually stratified by height; however, for our 2D WS$_2$ system, since the atoms are split into the discrete layers, we stratify by cluster instead, as described below.

The detached clusters above the surface are identified and removed at mid-impact, at the end of the impact, and after cooling. Clusters are detected by building a periodic-image-aware neighbor graph with a fixed cutoff distance and extracting its connected components. A cluster is determined to be removed from the surface when it meets all four of the following conditions: (a) it does not contain W atoms, (b) it has fewer than five atoms, (c) its center of mass sits at least $3.0~\text{\AA}$ above the highest W atom of the slab, and (d) it is moving away from the surface. The first two conditions avoid unphysical removal of W-containing clusters and restrict removal to small, W-free fragments such as O, O$_2$, S, S$_2$, SO, SO$_2$, etc. The last two confirm that a cluster is clearly separate from the slab and is moving away from the surface. We repeat the impact-cooling cycle until the surface following the impact reaches steady-state.

\subsection{Fine-tuning methods}
\label{sec:methods:ft}
\begin{figure}
    \centering
    \includegraphics[width=\linewidth]{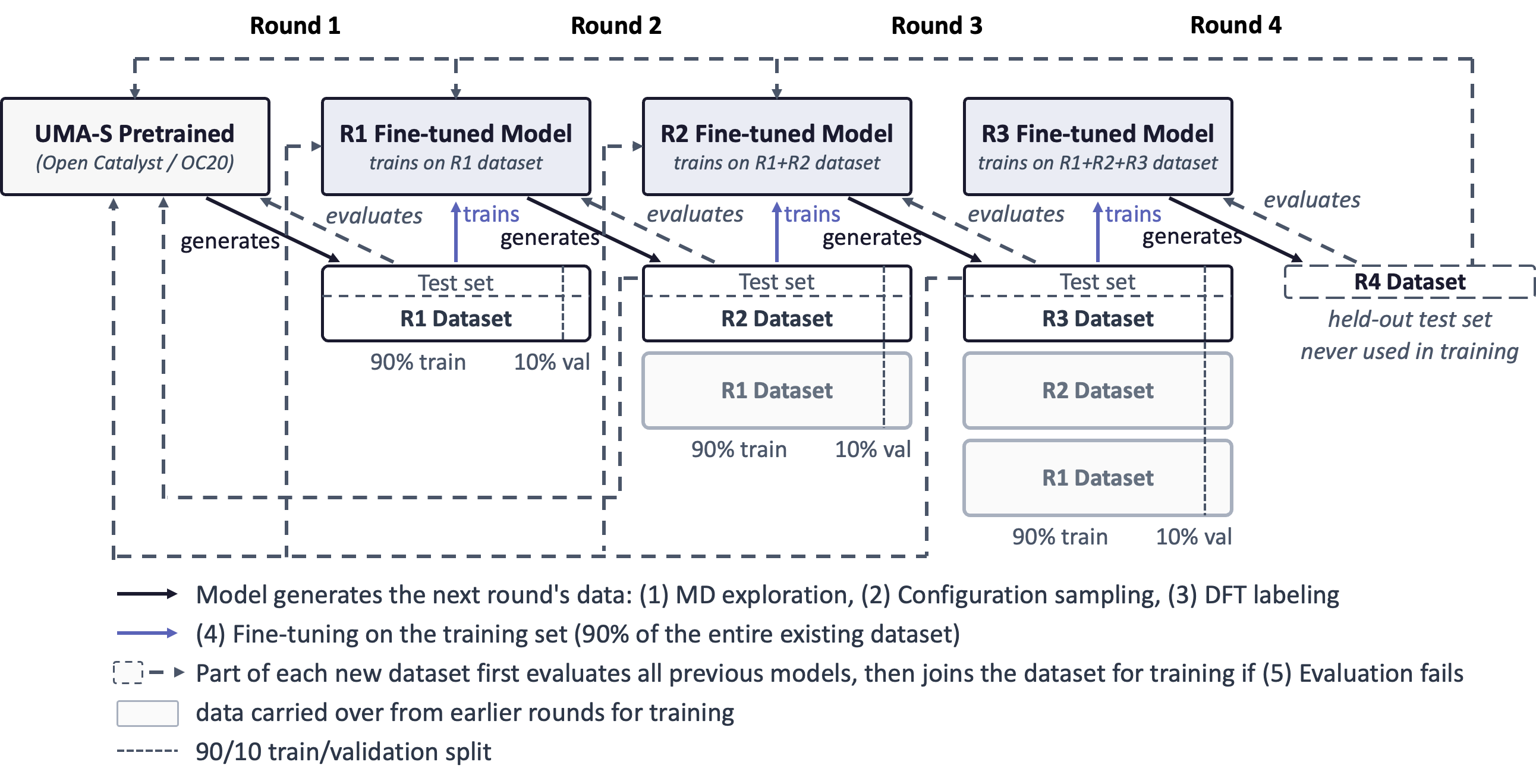}
    \caption{Round-by-round construction of the training and test sets in the iterative fine-tuning loop. Starting from the pretrained UMA model with the OC20 task~\cite{wood_uma_2026}, each round's model generates the next round's data through (1)~MD exploration; (2)~configuration sampling; (3)~DFT labeling (black arrows); (4)~fine-tuning (blue arrows); and (5)~evaluation (grey dashed arrows). See text for details.}
    \label{fig:finetuning_pipeline}
\end{figure}

A single round of fine-tuning on the dataset explored by the pretrained model is usually insufficient when the target chemistry is absent from the training set from the pretraining step~\cite{qi_robust_2024,liu_fine-tuning_2026}. Therefore, we adopt an iterative loop as shown in Fig.~\ref{fig:finetuning_pipeline}. Each iteration starts from the model of the previous iteration, or from the pretrained UMA-S in the first round. We run initial MD simulations to explore the W--S--O configuration space, choose a maximally diverse subset of the explored configurations, label them with DFT, and use those labels to update the model. The updated model is then used to run a new round of MD exploration, and newly labeled DFT configurations become the test set, which otherwise would be a portion of training sets in the next round. We call the loop `converged' once the model is sufficiently accurate on this test set and the predicted observables, namely the amounts of chemisorbed S and O in the simulation box as a function of ion dose, converge. If convergence is not reached, more configurations are labeled in addition to the previously labeled test set. The cumulative labeled data are used for the next iteration's training set, and the cycle continues.

\paragraph*{1. MD exploration}
At each iteration, the new configurations are generated by running MD simulations of O$^+$ and O$_2^+$ using the protocol described in Sec.~\ref{sec:methods:sim}. We use two cell geometries: a uniform $3\!\times\!3\!\times\!2$ slab, and a stepped two-layer slab in which a $3\!\times\!4$ top layer sits on a $4\!\times\!4$ bottom layer. The two geometries together cover both flat-surface chemistry and step-edge chemistry. After each run we strip the immobile bottom layer from both geometries, leaving only mobile atoms in the training configurations.

\begin{figure}
    \centering
    \includegraphics[width=\linewidth]{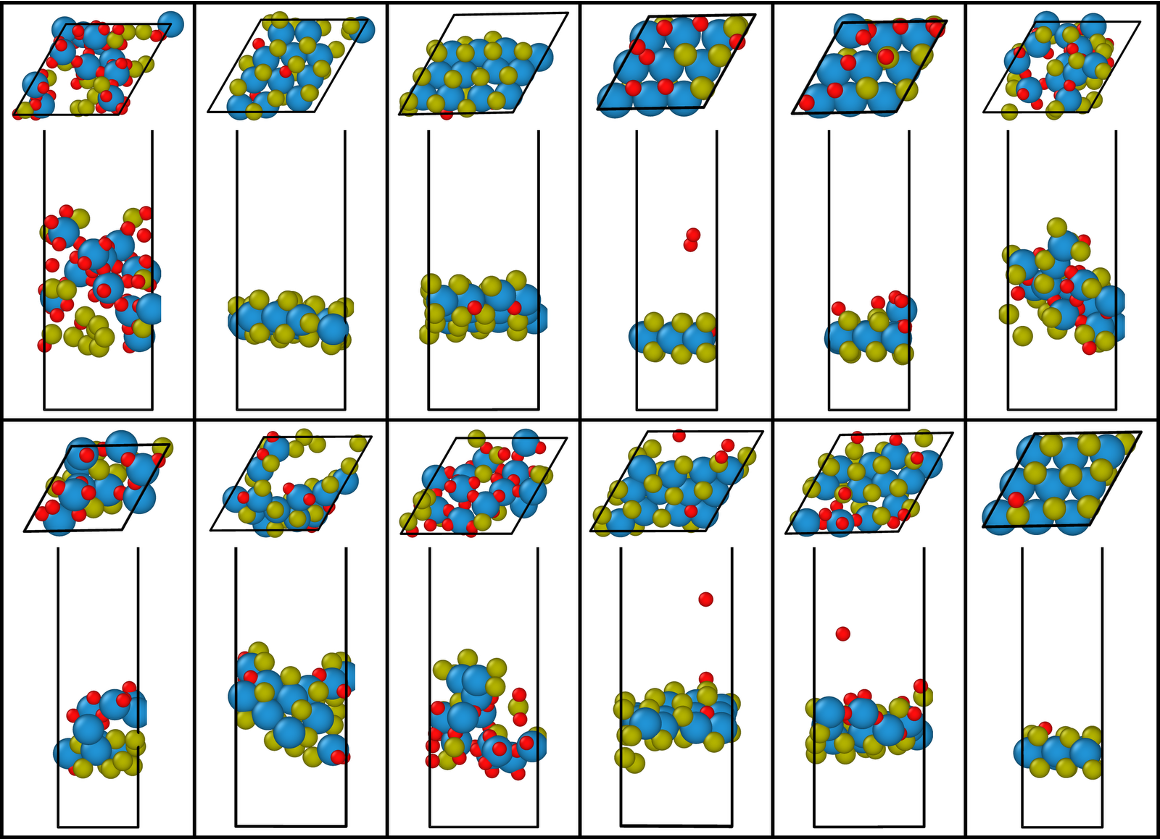}
    \caption{Representative configurations from the iterative fine-tuning training set are displayed. Twelve W--S--O configurations are drawn at random from the training set assembled by the SOAP+PCA+FPS protocol of Configuration Sampling for visualization. Each cell shows the top view above the side view. The immobile bottom layer has been stripped from every configuration as described in step~1. The training set includes both flat-surface and step-edge geometries at varying degrees of oxidation. W atoms are shown in blue, S in yellow, and O in red. Rendered with OVITO~\cite{stukowski2010visualization}.}
    \label{fig:training_set}
\end{figure}

\paragraph*{2. Configuration sampling}
This stage builds a training set that spans a wide range of local atomic environments. We treat every configuration the model explores as equally important, no matter how often it shows up in the simulation. We sample by combining a SOAP descriptor~\cite{bartok_representing_2013}, principal component analysis (PCA)~\cite{abdi2010principal}, and farthest point sampling (FPS)~\cite{imbalzano_automatic_2018}. SOAP encodes each atom's local environment, including neighbor identities and spatial arrangement of those neighbors within a cutoff radius, as a smooth, rotationally invariant fingerprint. We use the \texttt{dscribe}~\cite{himanen2020dscribe} implementation with W, S and O as species, a cutoff radius of $6.0$~\AA{}, and $n_{\mathrm{max}}=9$ radial and $l_{\mathrm{max}}=9$ angular basis functions, with the remaining parameters left at their defaults. SOAP is computed with inner averaging, so a single vector represents each configuration. Averaging is acceptable here because the immobile bottom layer has already been removed in step~1, leaving a single reactive layer with no additional layer beneath it. The oxidation is therefore not stratified in depth, and so the averaging reflects the changes. We compute SOAP for every candidate configuration, use PCA to cut the feature dimensionality to 50 principal components (including about 99\% of the explained variance), and run FPS in that reduced dimension space to pick the $N$ configurations that are mutually farthest apart in local-environment space. The SOAP, PCA and FPS steps are repeated afresh on each round's new pool, rather than being seeded with the labeled set from the previous rounds. Fig.~\ref{fig:training_set} shows representative configurations from the resulting training set.

\paragraph*{3. DFT labeling}
We label the sampled configurations with energy, force, and stress targets from periodic DFT in Quantum ESPRESSO (version 7)~\cite{giannozzi2009quantum,giannozzi2017advanced}. The calculations use the PBE exchange--correlation functional~\cite{perdew1996generalized} with spin polarization and Grimme's D3 van der Waals correction~\cite{grimme2010consistent}. On-site correlation on the W 5d manifold is treated with a Hubbard U of 2.87 eV~\cite{bui_first-principles_2015, dudarev_electron-energy-loss_1998} with orthogonalized-atomic-orbital projectors~\cite{timrov2020pulay}. The pseudopotentials are taken from the Standard Solid-State Pseudopotentials (SSSP) library (v1.3.0, PBE, efficiency)~\cite{prandini2018precision}: ultrasoft GBRV potentials for W and S~\cite{garrity2014pseudopotentials} and a projector augmented-wave (PAW) potential for O~\cite{dal2014pseudopotentials}. The plane-wave basis uses kinetic-energy cutoffs of 55~Ry (wavefunctions) and 430~Ry (charge density), and the Brillouin zone is sampled on a $\Gamma$-centered $2\!\times\!2\!\times\!1$ $k$-point mesh.

We standardize energy, force, and stress labels following the protocol described in Appendix~A.6 of Wood et al.~\cite{wood_uma_2026}. For the energy reference, we compute per-element isolated-atom DFT energies in a large cell at the same level of theory and combine them with tabulated heats of formation~\cite{mentel2021mendeleev}; this reference multiplied by the number of each element is subtracted from the DFT energy of each configuration. We apply a linear-regression to the referenced energies and scale every target (energy, forces, stress) by the force root-mean-square (RMS). For the regression coefficients and the force-RMS value, we reused the values that are computed for OC20, so the training set for the fine-tuning is on the same standard as that of OC20.

\paragraph*{4. Fine-tuning}
We use a 90/10 training-to-validation split on the labeled configurations. The fine-tuning runs in PyTorch~\cite{paszke2019pytorch} with custom code utilizing Wood et al.~\cite{wood_uma_2026} and Liu et al.~\cite{liu_fine-tuning_2026}. We train each iteration for 55 epochs under a cosine learning-rate scheduler with a 5-epoch linear warmup~\cite{loshchilov2016sgdr} and use exponential moving average (EMA) of the model weights. The training and validation loss are computed as follows:
\begin{equation}
\mathcal{L} = w_{E}\,\frac{1}{B}\sum_{i=1}^{B}
   \mathrm{MSE}\!\left(\frac{\hat{E}_{i}}{N_{i}},\,\frac{E_{i}}{N_{i}}\right)
+ w_{F}\,\frac{1}{B}\sum_{i=1}^{B}
   \mathrm{Huber}_{\delta}\!\left(\hat{F}_{i},\,F_{i}\right)
+ w_{S}\,\frac{1}{B}\sum_{i=1}^{B}
   \mathrm{MSE}\!\left(\hat{\sigma}_{i},\,\sigma_{i}\right),
\label{eq:fine-tuning-loss}
\end{equation}
where $B$ is the batch size, $N_{i}$ is the number of atoms in configuration $i$, $\hat{E}_{i}$, $\hat{F}_{i}$, and $\hat{\sigma}_{i}$ are the model predictions of the total energy, force components, and stress components for configuration $i$, respectively. $E_{i}$, $F_{i}$, and $\sigma_{i}$ are the corresponding DFT targets, and $w_{E}$, $w_{F}$, $w_{S}$ are the loss weights for the three channels. Notice that Huber loss with threshold $\delta$ is used for the force channel rather than an MSE, which keeps high-magnitude force outliers from dominating. We set $\delta=2$~eV/\AA{}, far above the typical force error, so the Huber loss acts only on the rare extreme residuals. Both the predictions and targets live in the force-RMS-normalized space as described above. The model is trained in this normalized space, and the FairChem calculator converts the output back to denormalized space at the time of inference by multiplying energy, force, and stress by the force-RMS. 

In this work, we test three fine-tuning strategies. Full fine-tuning updates all of the model parameters; layer freezing updates only the selected layers; and Low-Rank Adaptation (LoRA)~\cite{hu_lora_2021} freezes the per-expert MoLE blocks, which hold the majority of the model's parameters, and confines the update to the rank $r$ adapters of the frozen matrices along with the unfrozen parameters. Over the iterative refinement loop as depicted in Fig.~\ref{fig:finetuning_pipeline}, layer freezing is the primary strategy considered, meaning that R1, R2, and R3 fine-tuned models discussed in Sec.~\ref{sec:results} are trained with the layer freezing technique. Full fine-tuning and LoRA strategies are then used to compare with the R3 layer-frozen model to assess how the fine-tuning strategy affects the model performance, as in Sec.~\ref{app:results-finetune_methods}. None of the three strategies alters the graph construction. The neighbor-list cutoff, the neighbor-list settings, and the number of message-passing layers are unchanged from \texttt{uma-s-1p1}. For the LoRA strategy, we freeze the router until epoch 10. The reasoning for this approach is given in Sec.~\ref{app:router-freeze}. For each round and strategy, hyperparameter tuning was performed on the strategy-specific options (which layer(s) to freeze, LoRA rank, etc.), the loss weights $w_{E}$ and $w_{F}$, and the maximum learning rate. The model with the lowest validation loss is saved for analysis. The final model reported in Sec.~\ref{sec:results} uses $w_{E}=5$, $w_{F}=60$, and $w_{S}=10$.
 
\paragraph*{5. Evaluation}
At the end of each iteration, we evaluate the fine-tuned model on the test set which we acquire from the next round's sampled configurations as illustrated in Fig.~\ref{fig:finetuning_pipeline}. Because these configurations come from MD trajectories run with the current iteration's model, no model in the loop has seen them during training. The test set therefore measures how well the model generalizes to the configurations it has not seen during the training. It also places the R1, R2, \ldots, R$N$ fine-tuned models on equal terms for a round-to-round comparison of fine-tuning performance. As a minimum requirement, we enforce that the model's energy and force MAEs fall below 5~meV/atom and 0.1~eV/\AA{}~\cite{ko_recent_2023}. Meeting this standard is necessary but not sufficient: further criteria will be discussed in Sec.~\ref{sec:results}.

\section{Fine-tuning results and discussion}
\label{sec:results}
\begin{figure}
    \centering
    \includegraphics[width=\linewidth]{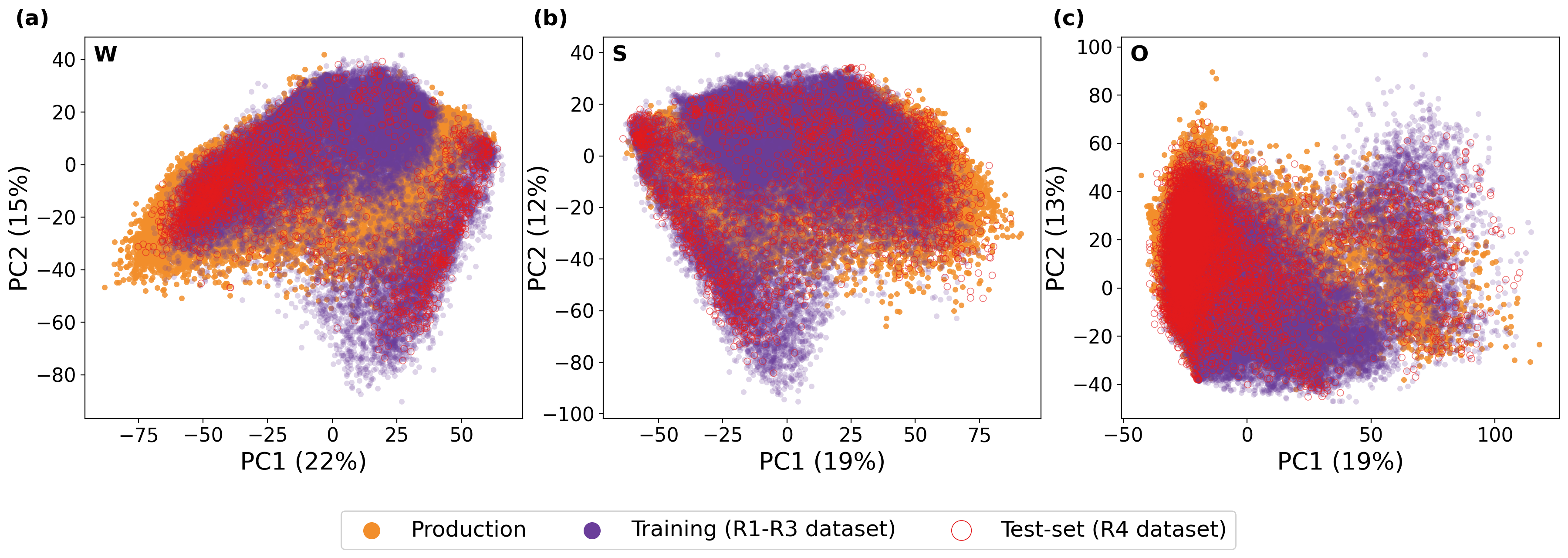}
    \caption{Coverage of the production-run local environment manifold by the training set across R1--R3 dataset and test set (R4 dataset) is displayed by projecting into a two SOAP+PCA principal components space for each element (a)~W, (b)~S, (c)~O. Every point is the SOAP fingerprint of a single atom in a single configuration, not a configuration-averaged descriptor. The selection itself was made on the configuration-averaged descriptor described in Sec.~\ref{sec:methods:ft}~(step~2), whereas this figure resolves the individual atomic environments that the selected configurations contain. Notice that this 2-component projection is coarser than the 50-component space used for the FPS sampling as described in Sec.~\ref{sec:methods:ft}~(step~2) for visualization. The orange points are SOAP fingerprints from the production-scale 15 eV O$^+$/O$_2^+$ bombardment-MD trajectories of Sec.~\ref{sec:methods:sim}, which is run with the R3 fine-tuned model and sampled every 10 impacts. The purple and the red-open points are the SOAP fingerprints from the training set across R1--R3 dataset and test set (R4 dataset), respectively. The fraction of total variance captured by each principal component is displayed on the axes.}
    \label{fig:training_production}
\end{figure}

The iterative fine-tuning procedure of Sec.~\ref{sec:methods:ft} must yield a fine-tuned UMA model that is accurate during the MD `production' runs, with a larger cell and more atoms than used in the training. The results plotted in Fig.~\ref{fig:training_production} can help determine this by examining the degree to which the training set spans the configuration space. For each element (W, S, and O), we project the SOAP fingerprints of every atom seen in R3 production trajectories into a 2-principal component space and compare with that of the training sets used for training R3 fine-tuned model (R1--R3) and the test sets that are generated by R3 fine-tuned model. The figure shows that a small population of production-run local environments lies outside the training envelope. This is clear in both W-panel and S-panel where some SOAP fingerprint points from the production run correspond to PC1~$\lesssim -50$ (W-panel) and PC1~$\gtrsim 60$ (S-panel). Some of the gap, however, is covered by the R4 test set. Because the R4 dataset is sampled from trajectories generated with the same R3 fine-tuned model used for the production-scale simulations, its local environments are consistent with those of the production run, with no evidence of a coverage gap. The training sets, by contrast, are generated by the pretrained, R1, and R2 models, which is why they leave the small gap noted above (Sec.~\ref{sec:methods:ft}, step~5).

These observations support a few elements of the chosen methodology. First, a gap between the accumulated training data and the trajectories generated by the newly fine-tuned model shows that we need some form of refinement loop when training the MLIP. Second, the assessment of an MLIP should be from a true held-out test set, which the model has not seen during the training. If the test set were chosen from the same chemical environment, the evaluation might exclude some of the out-of-distribution configurations that could be meaningful for the dynamics. Lastly, the gap between the SOAP fingerprints of the production-scale run and the test set appears minimal. The projection shows no sign of disagreement between the smaller-cell and bigger-cell environment. This supports the decision to use smaller cells for the training, which are computationally more efficient.  We examine this point further in Sec.~\ref{app:OOD_performance}, where we evaluate the model's accuracy on configurations generated from simulations with a slightly larger cell. The same projection also speaks to whether FPS, a diversity-based selection method, neglects commonly visited geometries in favor of distribution extremes. The training points cover most of the production manifold, including the dense interior where the ordinary environments lie as well as the edges.

\subsection{Fine-tuning performance over rounds}
\label{sec:results-finetune_rounds}
\begin{figure}
    \centering
    \includegraphics[width=\linewidth]{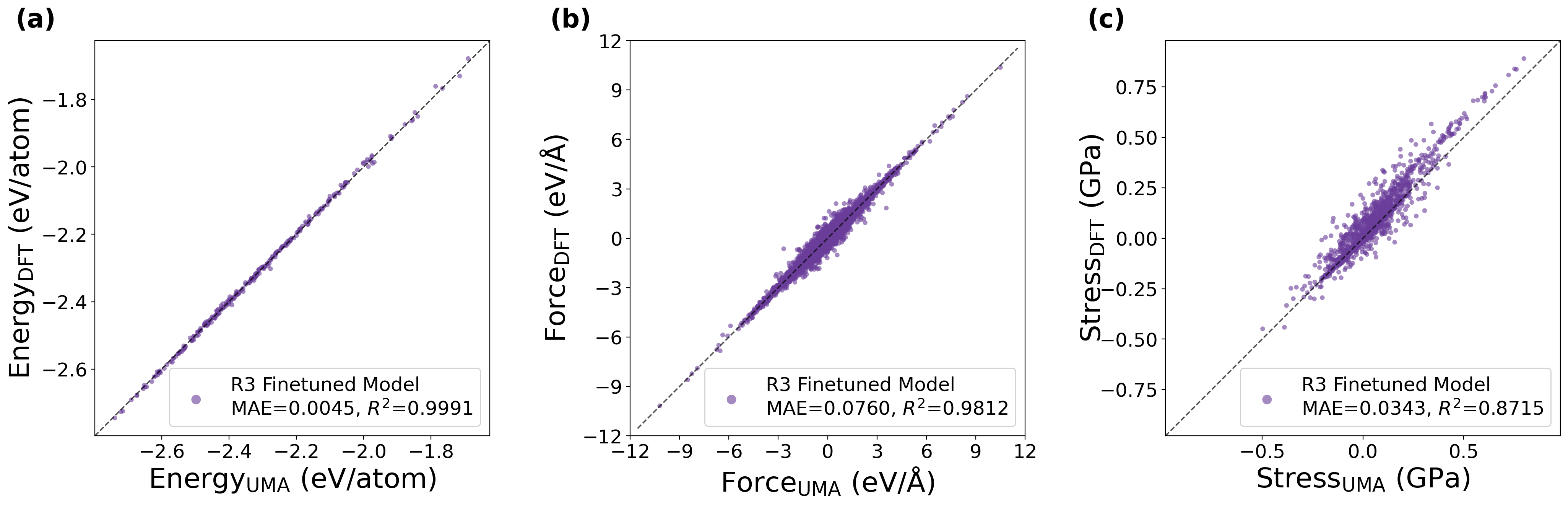}
    \caption{Parity plot of the R3 fine-tuned model vs. the PBE+D3+$U$+spin DFT reference on the R4 test set is displayed for (a)~energy (MAE${=}4.5\times 10^{-3}$~eV/atom, $R^{2}{=}0.999$), (b)~forces (MAE${=}0.076$~eV/\AA, $R^{2}{=}0.981$), and (c)~stress (MAE${=}0.034$~GPa, $R^{2}{=}0.872$). Points that deviate from the dashed line (y$=$x) show greater error between the fine-tuned model and DFT. Energies are total energy per atom.}
    \label{fig:r3_finetuned_parity}
\end{figure}

The R3 fine-tuned model appears to accurately match the PBE+D3+$U$+spin DFT reference on the R4 test set as shown in Fig.~\ref{fig:r3_finetuned_parity}. The energy MAE is $4.5\times 10^{-3}$~eV/atom ($R^{2} = 0.999$), the force MAE is $0.076$~eV/\AA{} ($R^{2} = 0.981$), and stress MAE is $0.034$~GPa ($R^{2} = 0.872$). We use force MAE as the primary target because the MD simulation is driven by forces, but stress is retained as a secondary loss term so that the atoms in the cell are not experiencing inaccurate stress.

\begin{figure}
    \centering
    \includegraphics[width=0.8\linewidth]{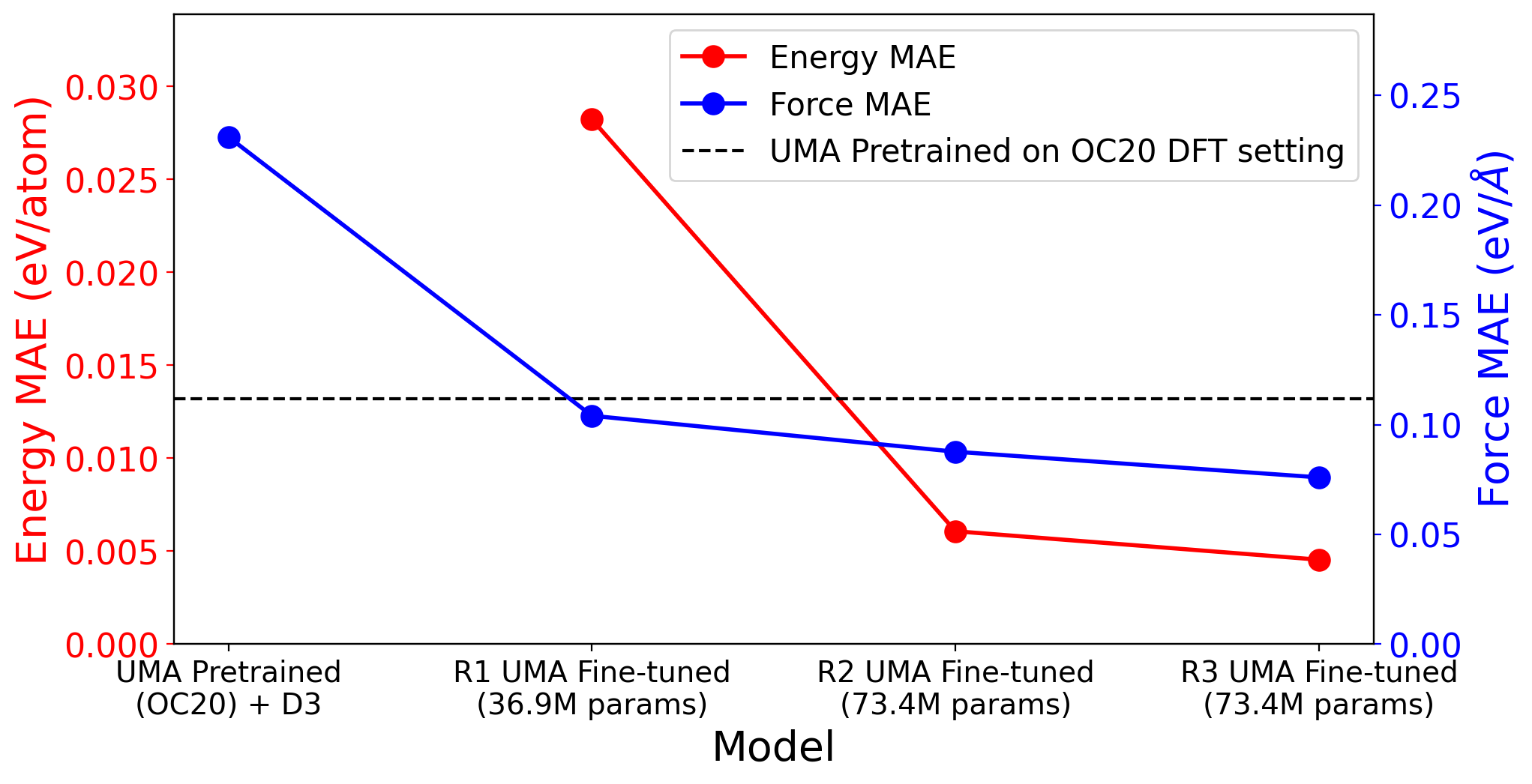}
    \caption{Round-over-round energy MAE (red, left axis, eV/atom) and force MAE (blue, right axis, eV/\AA{}) on the R4 test set are evaluated. The values in the parentheses indicate the number of trainable-parameters for each model during the fine-tuning. The dashed horizontal line ("OC20 setting") marks the energy and force MAE of the pretrained UMA-S model evaluated on the test set under its native OC20 task setting (see Sec.~\ref{app:oc20_eval} for how this evaluation is performed). The pretrained baseline is omitted for energy MAE evaluation because its energy reference is different from that of fine-tuned models so they are not directly comparable. Energies are total energy per atom.}
    \label{fig:different_rounds_performance}
\end{figure}

\begin{figure}
    \centering
    \includegraphics[width=\linewidth]{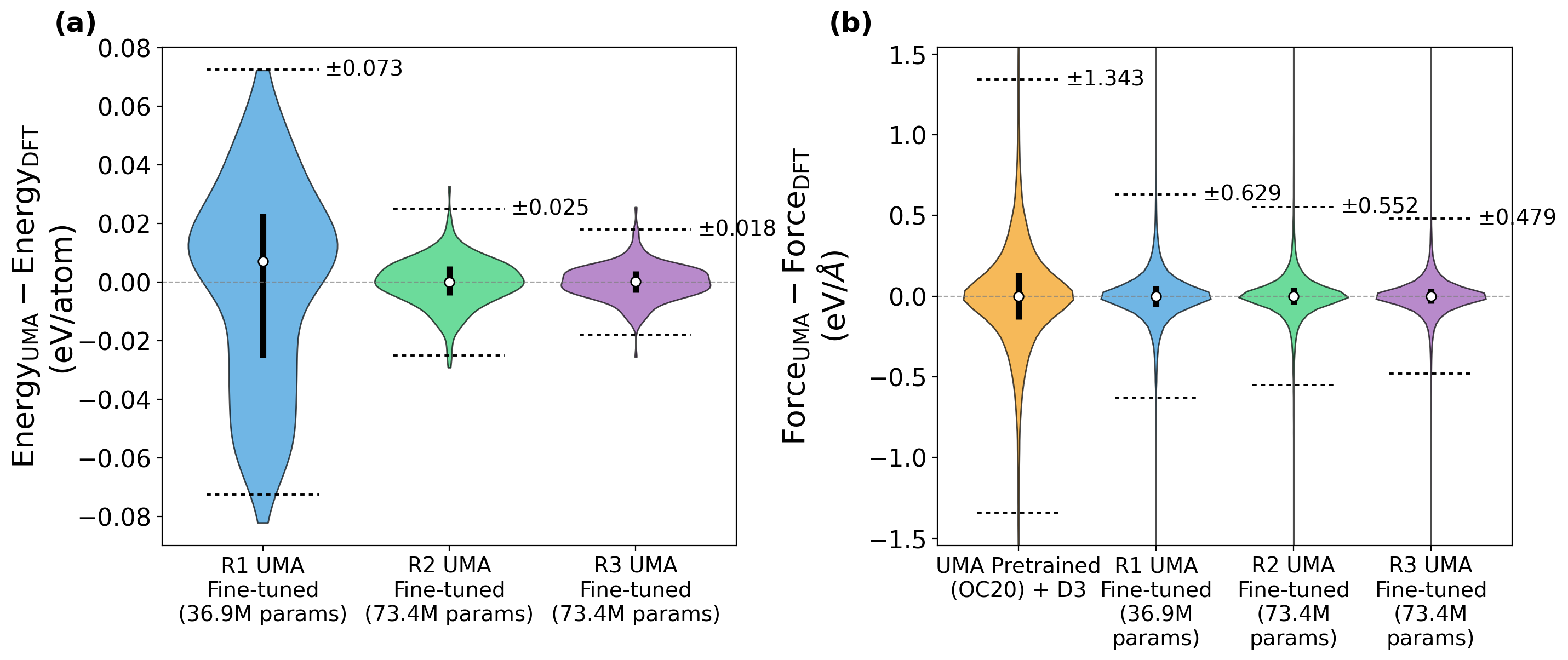}
    \caption{Violin plot of the error distribution (UMA $-$ DFT) on the R4 test set. (a)Energy error (eV/atom) for the R1, R2, and R3 fine-tuned models and (b)Force error (eV/\AA{}) for the pretrained UMA-S~(OC20)+D3 baseline and for the R1--R3 fine-tuned models. The pretrained baseline is omitted for energy MAE evaluation because its energy reference is different from that of fine-tuned models so they are not directly comparable. The white dot indicates the median, the inner black box marks the interquartile range, and the dashed line illustrates the 99\% inclusion bound of each distribution. The latter is annotated with the corresponding $\pm$ value. Energies are total energy per atom.}
    \label{fig:violin_plot}
\end{figure}

 Fig.~\ref{fig:different_rounds_performance} shows the performance of the fine-tuning over different rounds on both energy and force MAEs. Notice that energy MAE of the pretrained model is omitted, since the energy reference of the pretrained model under OC20 task differs from that of PBE+D3+$U$+spin used for labeling the fine-tuning data and so they are not directly comparable (same for Fig.~\ref{fig:violin_plot}). Instead, the pretrained model's energy and force MAE are evaluated on the test set under its native OC20 task setting (see Sec.~\ref{app:oc20_eval}), giving $0.013$~eV/atom and $0.112$~eV/\AA{} for energy and force MAE. Scored against the PBE+D3+$U$+spin reference, the pretrained force MAE is evaluated as $\approx 0.23$~eV/\AA{}, which reflects the functional mismatch. By R3, fine-tuning brings the force MAE down to $0.076$~eV/\AA{}, below even the native OC20 baseline by roughly a factor of 1.5; the energy MAE drops from $2.8\times 10^{-2}$ to $4.5\times 10^{-3}$~eV/atom between the R1 and R3 fine-tuned models. The largest single round gain for force was between the pretrained model and R1 fine-tuned model and for energy was between the R1 and R2 fine-tuned models, after which the gains shrank. Such diminishing-return behavior motivates stopping at round 3, since further rounds would be expected to reduce the MAE only marginally.

The round-over-round MAE as illustrated in Fig.~\ref{fig:different_rounds_performance} is useful, but because it averages over the whole error distribution, it can sometimes mask `heavy-tailed' behavior that can become important for MD stability. Fig.~\ref{fig:violin_plot} reports the distribution of the energy and force difference between UMA and DFT at each round. In Fig.~\ref{fig:violin_plot}, the inner box is used to indicate the interquartile range, the white dot is the median, and the dashed line defines the 99\% inclusion bound of each distribution. Both medians and interquartile ranges are centered around 0, with the exception of the R1 UMA fine-tuned model. With the least training data and three frozen layers rather than two, it could be that the R1 model conflates two different levels of DFT theory: OC20 setting (RPBE) and PBE+D3+$U$+spin. However, such asymmetry is gone by round 2, where both energy and force violins are symmetric and the predictions are centered around zero. Moreover, the 99\% inclusion bounds shrink over the rounds of fine-tuning for both energy and force channel. It shrinks from $\pm 0.073$~eV/atom at R1 to $\pm 0.018$ at R3 for energy and from $\pm 1.343$~eV/\AA{} for the pretrained baseline to $\pm 0.479$ at R3 for force. When considering errors associated with MD stability, this tightening of the outlier tails is just as important as reducing MAE values since a single large force residual can destabilize a trajectory that might otherwise be well-behaved.

\begin{figure}
    \centering
    \includegraphics[width=\linewidth]{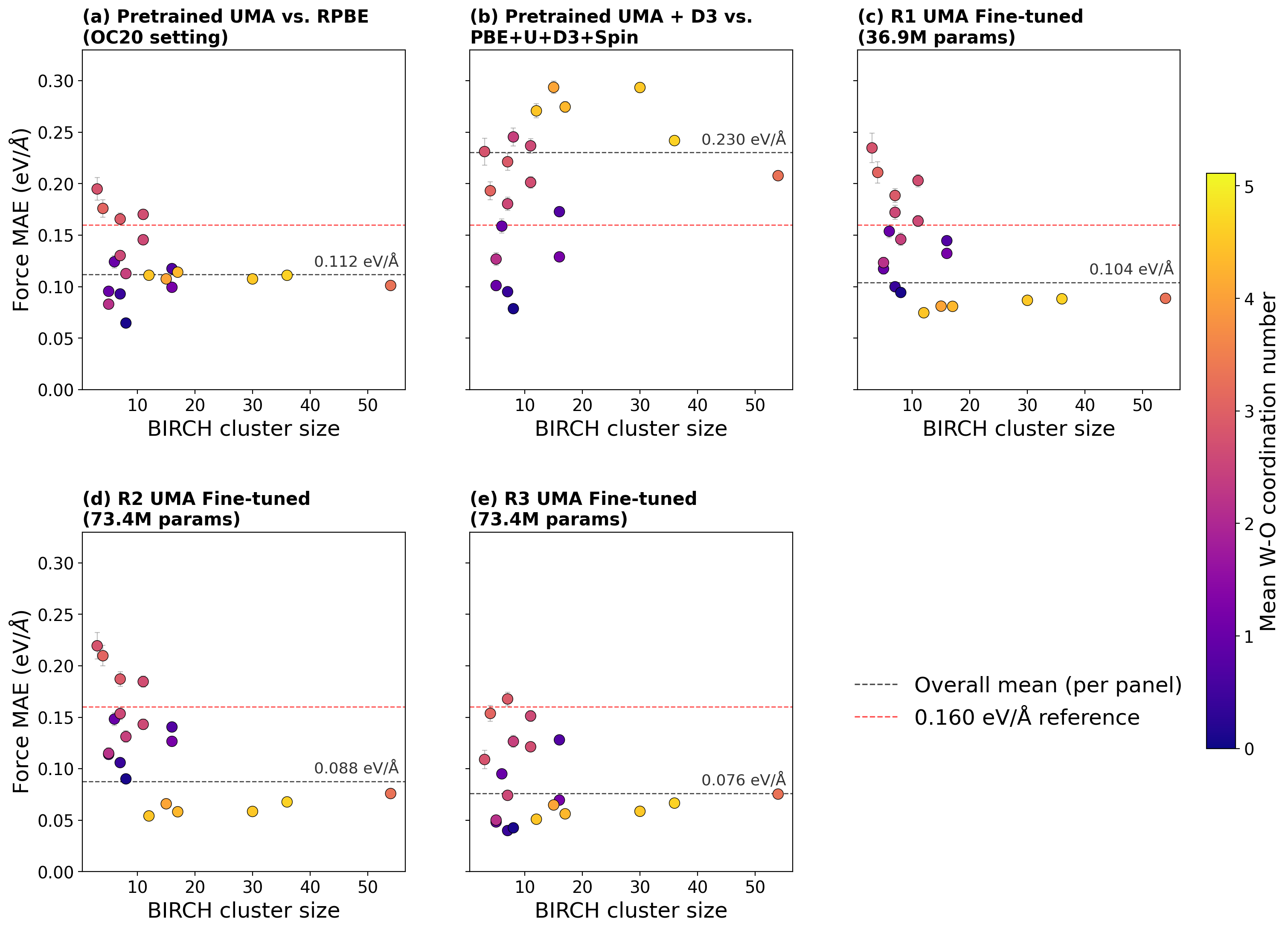}
    \caption{Local-environment-resolved force MAE on the R4 test set across (a)~the pretrained UMA-S model evaluated under the OC20 task setting, (b)~the pretrained UMA-S\,+\,D3 model, (c)~R1 fine-tuned model, (d)~R2 fine-tuned model, and (e)~R3 fine-tuned model is displayed. Panels (b) through (e) are all evaluated against the same PBE+D3+$U$+spin reference used to label the fine-tuning data, whereas panel (a) is evaluated under the OC20 task setting (see Sec.~\ref{app:oc20_eval}). The test set configurations are grouped into 20 BIRCH clusters from the SOAP+PCA space defined in Sec.~\ref{sec:methods:ft}~(step~2). Each marker is for each cluster, and they are plotted based on the number of configurations within the cluster and mean force MAE of the cluster, with the spread of force MAE across the cluster displayed with error bars. They are also colored based on their mean W-O coordination number of the cluster's atoms. The black dashed line is the per-panel mean force MAE (0.112, 0.230, 0.104, 0.088, and 0.076~eV/\AA{} in panels a--e), and the red dashed line at 0.160~eV/\AA{} is a diagnostic threshold for per-cluster force MAE. The number of trainable parameters is indicated in the parentheses for each fine-tuned model.}
    \label{fig:birch_finetuning_rounds}
\end{figure}

The local-environment resolved BIRCH~\cite{zhang1997birch} analysis in Fig.~\ref{fig:birch_finetuning_rounds} can provide a complementary picture of the source of potential errors. The test set configurations are grouped into 20 clusters from the SOAP+PCA latent space, similar to the configuration sampling step described in Sec.~\ref{sec:methods:ft}. Force MAE values are averaged for each cluster. The clusters are colored based on their mean W-O coordination number. Against the PBE+D3+$U$+spin reference (panel (b)), the pretrained model performs worst on the high mean W-O coordination number clusters and better with lower mean W-O coordination number. However, this appears to be an artifact of the reference mismatch, since, under its native OC20 setting (panel (a)), the high-coordination clusters are just around the average. Fine-tuning then reduces the high coordination errors just within a single round, whereas errors associated with mid-W-O coordination clusters (roughly 2--4) drop more gradually over the rounds. In any case, by round 3 the per-cluster force MAE is near or below the 0.160~eV/\AA{} mark, which is likely to be sufficiently accurate for production-scale MD simulation.

Overall, Figs.~\ref{fig:r3_finetuned_parity}--\ref{fig:birch_finetuning_rounds} suggest the following points about the fine-tuning loop. The R3 model reaches energy and force MAE below $\sim 5\times 10^{-3}$~eV/atom and $\sim 0.08$~eV/\AA{}, respectively, on a true held-out set. Further, the round-over-round improvement comes from gradually reducing the errors associated with mid-coordination or partially-oxidized W local environments. The results above use a single fine-tuning strategy, i.e. layer freezing. Whether the converged accuracy depends on this choice, and how the parameter-efficient LoRA variants compare, is examined in Sec.~\ref{app:results-finetune_methods}.

\subsection{Plasma-surface dynamics}
\label{sec:results-md}
The plasma-surface observables of interest in this study of oxygen plasma exposed to WS$_2$ films include the removal of chemisorbed S and the uptake of chemisorbed O in the top layer. It is known that oxygen plasma exposure to WS$_2$ layers, under the conditions we examine here, results in complete removal of S and chemisorption of O to W to form a layer of WO$_3$~\cite{DonnellyPC}. Minimal or no W is expected to etch under the present conditions.

Though the R3  model reaches an energy MAE of $4.5\times 10^{-3}$~eV/atom and force MAE of $0.076$~eV/\AA{}, reaching those targets is not sufficient: the final model also has to generate stable, physically reasonable trajectories under the production-scale MD simulations of O$^+$ and O$_2^+$ described in Sec.~\ref{sec:methods:sim}. Just as the test set MAEs level off over the fine-tuning rounds, we expect the physical observables obtained from the production-scale MD simulations to converge over the rounds once the loop has converged. 

To test for this convergence, we apply the pretrained UMA-S model (with D3 correction included) and R1, R2, and R3 fine-tuned models to the production-scale MD simulation. As described in Sec.~\ref{sec:methods:sim}, a $6\!\times\!6\!\times\!3$ 2H-WS$_2$ slab is bombarded with O$^+$ or O$_2^+$ ions at 15~eV, and we track the chemisorbed-S and chemisorbed-O coverages. Those observables are computed as a function of ion dose, i.e., the number of ion impacts normalized by the areal density. Ideally, we will see that the predicted coverages as a function of ion dose will converge as the fine-tuning rounds progress.

\begin{figure}
    \centering
    \includegraphics[width=\linewidth]{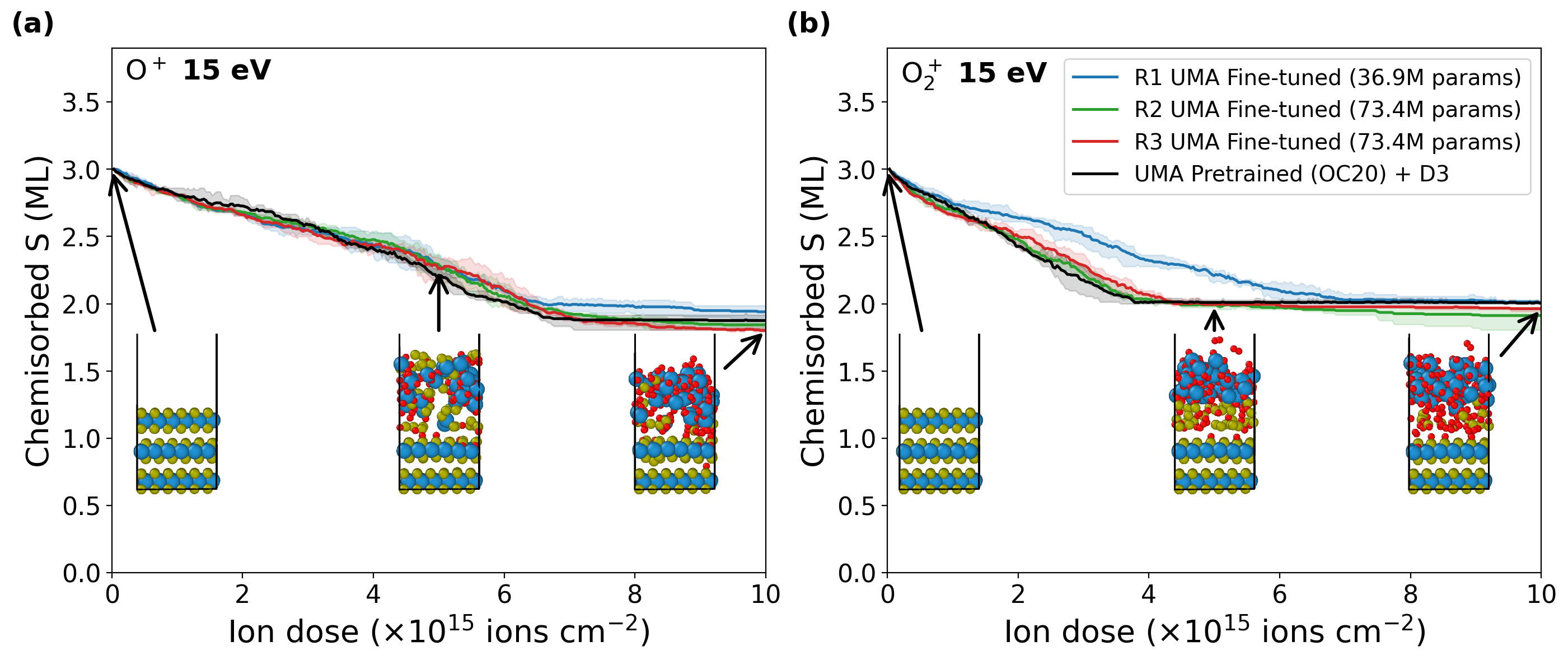}
    \caption{Remaining chemisorbed S in the simulation box is plotted as a function of ion dose for pretrained UMA-S model (with D3 included) and R1, R2, and R3 fine-tuned models during (a) 15 eV O$^+$ bombardment and (b) 15 eV O$_2^+$ bombardment. The chemisorbed S content is reported in terms of monolayers (ML), defined as $1\,\mathrm{ML} = 72$~S atoms, corresponding to two S atoms per W in  WS$_2$. The solid line represents the mean chemisorbed S over ion dose of the three different replicates, and the shaded band shows the minimum and maximum across different replicates. The snapshots from the R3 fine-tuned model for every $5\times 10^{15}$~ions/cm$^{2}$ are displayed, which are rendered with OVITO~\cite{stukowski2010visualization}.}
    \label{fig:chemisorbed_S_uptake}
\end{figure}

Fig.~\ref{fig:chemisorbed_S_uptake} plots chemisorbed S surface coverage as a function of ion dose, with the pristine slab starting at 3~ML of S. 1~ML of S is defined as 72 atoms, which corresponds to two S atoms per W in a single WS$_2$ layer. The sulfur atom was deemed chemisorbed if (a) it has W atoms within 2.85~\AA{} and (b) it has less than or equal to 1 oxygen neighbor within 1.75~\AA{}. These stipulations exclude weakly bound SO$_2$ that may be near a W atom but is not chemically bound to it. Fig.~\ref{fig:chemisorbed_S_uptake} shows that the chemisorbed-S content for both O$^+$ and O$_2^+$ impacts decreases monotonically with ion dose and levels off near 1.8--2.0~ML of S. This result means that S in the top layer is fully removed, with minimal damage or disruption to the layer below. Notably, the pretrained UMA-S model reproduces this profile closely, despite using neither spin polarization nor a Hubbard $U$ correction. Results from the R2 and R3 fine-tuned models trace essentially the same curve. The R1 model deviates slightly from this converged profile, consistent with the transient R1 offset noted earlier (Fig.~\ref{fig:violin_plot}). The saturation for 15~eV O$^+$ occurs at an ion dose of $\sim 7\times 10^{15}$~ions/cm$^{2}$ and saturation for 15~eV O$_2^+$ occurs at $\sim 4\times 10^{15}$~ions/cm$^{2}$ for the pretrained model and R2 and R3 fine-tuned models.

\begin{figure}
    \centering
    \includegraphics[width=\linewidth]{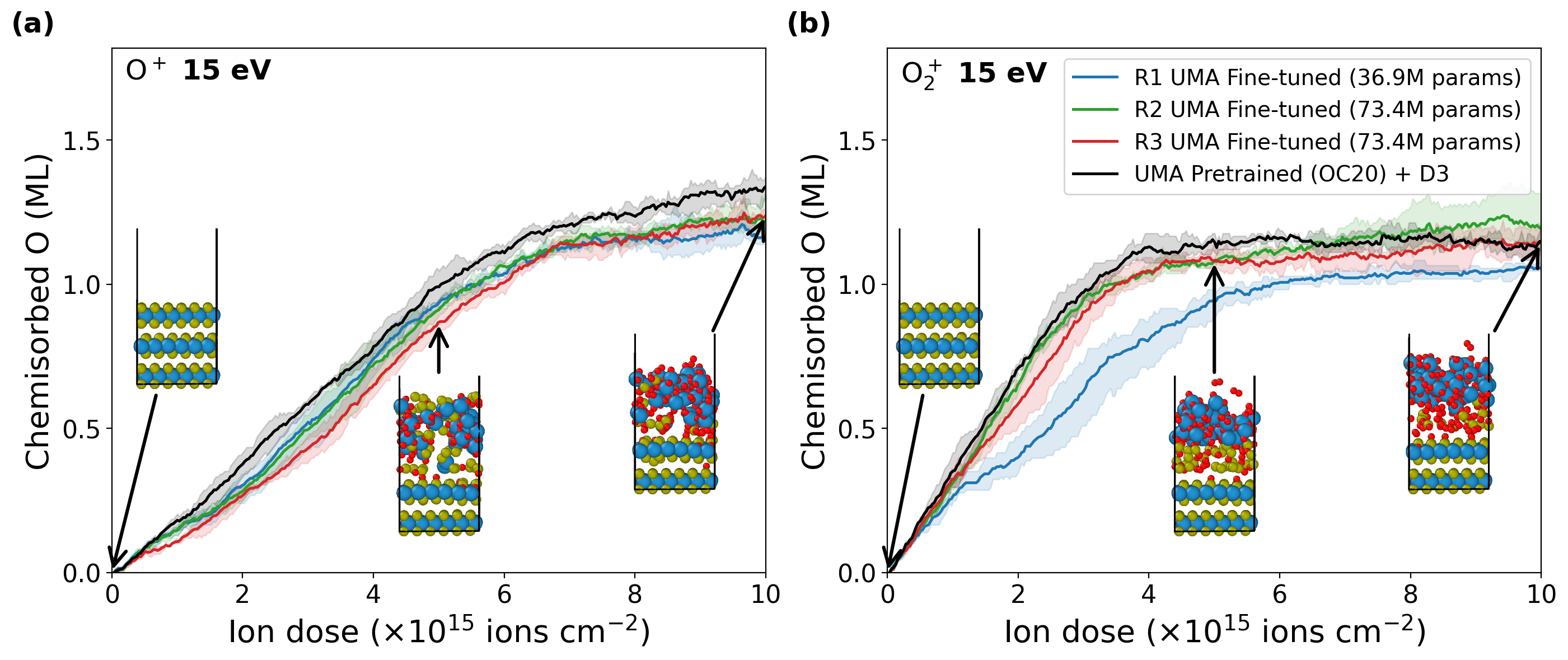}
    \caption{Chemisorbed O in the simulation box is plotted as a function of ion dose for pretrained UMA-S model (with D3 included) and R1, R2, and R3 fine-tuned models during (a) 15 eV O$^+$ bombardment and (b) 15 eV O$_2^+$ bombardment. The chemisorbed O content is reported in terms of monolayers (ML), defined as $1\,\mathrm{ML} = 108$~O atoms, corresponding to three O atoms per W in  WO$_3$. The solid line indicates the mean chemisorbed O over ion dose of the three different replicates, and the shaded band demonstrates the minimum and maximum across different replicates. The snapshots from R3 fine-tuned model for every $5\times 10^{15}$~ions/cm$^{2}$ are displayed, which are rendered with OVITO~\cite{stukowski2010visualization}.}
    \label{fig:chemisorbed_O_uptake}
\end{figure}

Similarly, Fig.~\ref{fig:chemisorbed_O_uptake} shows the chemisorbed O uptake as a function of ion dose. 1~ML of O was defined as 108 atoms, which corresponds to three O atoms per W as in WO$_3$. An oxygen atom was considered chemisorbed if (a) it has no O neighbor within 1.35~\AA{}, and either (b) it has a W atom within 2.3~\AA{} or (c) it is bonded to a chemisorbed S within 1.75~\AA{}. These conditions minimize the chance of classifying a weakly bound O as being chemisorbed to WS$_2$. Similar to the chemisorbed-S case, the chemisorbed-O uptake rises monotonically from zero and saturates at a higher dose. For O$^+$, the pretrained model and R1, R2, and R3 fine-tuned models all give essentially the same uptake trajectory. For O$_2^+$, fine-tuned models R2 and R3 fall on the same line, converging at a dose of $\sim 3.5\times 10^{15}$~ions/cm$^{2}$ above 1.0 ~ML of O, whereas R1 converges later at $\sim 5.5\times 10^{15}$~ions/cm$^{2}$, and saturates below the others, again reflecting the transient R1 offset in Fig.~\ref{fig:violin_plot}.

The coverage curves record what remains on the surface, so we also tracked the identity and amount of the sulfur that leaves it. SO and SO$_2$ are the largest etched products, and the rest is etched as S, S$_2$, S$_2$O, and SO$_3$. These counts include only the fragments removed by the detached-cluster criteria of Sec.~\ref{sec:methods:sim}, and physisorbed species stay on the surface. Although those criteria exclude W-containing clusters, no W atom detached from the surface in any run. The R2 and R3 fine-tuned models track each other closely here, so the loop has converged on this observable as well, whereas the pretrained model separates from both. Fine-tuning therefore does change the finer details of the dynamics, even where the coverage curves agree. The comparison is given in Sec.~\ref{app:etched_S}.

These results demonstrate that the pretrained model and the R2 and R3 fine-tuned models predict essentially the same chemisorbed S and O coverages, with R1 the only outlier, consistent with the R1 offset in Fig.~\ref{fig:violin_plot}. In other words, the chemisorbed coverages are captured by the pretrained UMA-S model, while fine-tuning improves accuracy and changes the etched-product distribution. In any case, the fine-tuning loop appears to have converged in both the test set MAE sense and the production-observable sense.
 
\section{Summary and concluding remarks}
\label{sec:conclusion}
We report the development and evaluation of a G-MLIP-based foundation model, employing UMA-S under the OC20 task as a pretrained model, for MD simulations of energetic (15 eV) oxygen ions (O$^+$ and O$_2^+$) impacting several layers of 2D TMD WS$_2$. Each round of the iterative loop ran cumulative ion-impact MD with the current model, selected a maximally diverse set of atomic configurations using SOAP and FPS, and labeled the selected configurations with spin polarized DFT at the PBE level with D3 dispersion and Hubbard $U$ corrections. The model was then re-trained on all configurations labeled up to that round, with a 90/10 training-to-validation split. The held-out test set was created by generating new configurations with the updated model. Since no fine-tuned models were trained on these configurations, the test set provides a fair comparison across all fine-tuning rounds. 

With three rounds of fine-tuning, the MAE for energy and force reduced to $4.5\times 10^{-3}$~eV/atom and $0.076$~eV/\AA{}, evaluated on the held-out test set. This is an improvement from the pretrained model even when it is evaluated against labels computed at OC20-consistent DFT settings (energy MAE of $0.013$~eV/atom and force MAE of $0.112$~eV/\AA{}), which do not have Hubbard $U$ correction and spin polarization that could be important to dynamics~\cite{bui_first-principles_2015}. The violin plots show that the distribution of the energy and force errors is symmetric and the 99\% inclusion bound has improved over the rounds. A local-environment-resolved BIRCH analysis showed that the biggest disagreement with the test set comes from the mean W-O coordination value between 2 and 4, where these force errors are reduced to around 0.16~eV/\AA{} after the third round of fine-tuning. Finally, MD predictions of the experimental observables of chemisorbed S removal and chemisorbed O uptake are compared across the pretrained model and the fine-tuned models. The pretrained UMA-S model reproduces these observables well, even without spin polarization and a Hubbard $U$ correction, and the fine-tuned predictions converge to the same result after three rounds. The etched sulfur species examined in Sec.~\ref{app:etched_S} are more discriminating. The R2 and R3 models show a good agreement there, while the pretrained model shows a different trend from both, so fine-tuning does change the finer details of the dynamics. A quantitative comparison with experiment will be presented in a subsequent paper.

\section*{Acknowledgments}
The authors gratefully acknowledge valuable discussions with Professor V.M. Donnelly and Dr. J. Mettler (University of Houston). This material is based upon work supported by the U.S. Department of Energy, Office of Science, Fusion Energy Sciences and Basic Energy Sciences, as part of the Extreme Lithography \& Materials Innovation Center (ELMIC), a Microelectronics Science Research Center (MSRC), under contract number No. DEAC02-09CH11466.
\section*{Data Availability Statement}
The weights for the R3 fine-tuned model are available on the Hugging Face Hub at \url{https://huggingface.co/RussellKwon/UMA_s1p1_oc20_fine-tuned_WSO} (DOI: \href{https://doi.org/10.57967/hf/9355}{10.57967/hf/9355}). The model is a derivative of UMA-S 1.1, so any redistribution must pass along the same FAIR Chemistry License (\S1.b.i), and any publication using it should acknowledge the use of UMA (\S1.b.ii).

\printbibliography

\appendix
 
\section*{Appendix}
\addcontentsline{toc}{section}{Appendix}
\renewcommand{\thefigure}{A\arabic{figure}}
\setcounter{figure}{0}
\renewcommand{\thesubsection}{A.\arabic{subsection}}
\setcounter{subsection}{0}

\subsection{Choice of the OC20 task over OMat}
\label{app:oc20-omat}
As noted in Sec.~\ref{sec:methods}, the pretrained UMA-S model was applied to the WS$_2$ system under the OC20 task rather than the OMat task. The choice was made by benchmarking the two tasks against ab-initio MD (AIMD) reference trajectories of 15~eV O$^+$ impacts on a $3\!\times\!3\!\times\!1$ WS$_2$ slab small enough to be tractable for AIMD. Three independent AIMD trajectories, each with a different random impact position, were run at the PBE+D3 level (with no spin polarization or Hubbard $U$ correction) with Quantum ESPRESSO~\cite{giannozzi2009quantum,giannozzi2017advanced}. The pretrained UMA-S model was then evaluated as a single-point calculator on every AIMD frame under each of the two tasks, with the D3 correction added to both so that all forces are compared at a consistent level of theory. The frame-by-frame force-parity comparisons against the AIMD reference are shown in Supplementary Movies~S1--S2.

Both task assignments track the AIMD reference accurately during the initial collision, namely the part of each trajectory in which the incoming O atom transfers its 15~eV of kinetic energy into the substrate. The two tasks then diverge from AIMD in different post-collision regimes. Under the OC20 task, the predicted forces lose fidelity once a sputtered S atom comes into bonding contact with another S atom and forms an S--S bond (Supplementary Movie~S1). This behavior is consistent with the OC20 training distribution, which is built around single-adsorbate-on-surface configurations~\cite{chanussot_open_2021} and therefore underrepresents S--S bonding which is due to disruption in the substrate lattice from the high energy impact. Under the OMat task, the predicted forces lose fidelity once the incoming O atom sits at an intermediate height of approximately 2--4~\AA{} above the substrate, namely in the regime where O is no longer in the gas phase but has not yet bonded into the lattice (Supplementary Movie~S2). This behavior is consistent with the OMat training distribution, which is dominated by equilibrium and near-equilibrium bulk-materials configurations~\cite{barros2026open} and therefore underrepresents the long-range adsorbate-surface interaction regime.
 
Both failure modes can in principle be removed by fine-tuning, but they are not equally easy starting points. The types of configurations that can be observed in plasma-surface interaction studies include an ion approaching, striking, and reacting with the surface, which are closer to the OC20 task that is constructed on the adsorbate-on-surface geometry. The OMat task, on the other hand, fails when the O$^+$ ion strikes the surface, disrupts the substrate, and reflects to a height of 2--4~\AA{}. This would be much more challenging than learning from OC20's failure mode, S--S bond formation due to high energy collision because using OMat as a baseline would require learning from the entire long-range adsorbate-surface interaction profile, which would require more learning. The OC20 task is therefore the closer starting point for the plasma-surface interaction studied here and was selected for the iterative fine-tuning loop of Sec.~\ref{sec:methods:ft}. This decision is reinforced by the production-scale results of Sec.~\ref{sec:results-md}: even without fine-tuning, the pretrained model under OC20 reproduces the target observables well. Its main shortcoming is the underrepresented S--S bonding regime, and this does not affect the observables we care about.

\subsection{Benchmarking the pretrained model under its native OC20 setting}
\label{app:oc20_eval}
Under the OC20 task, the pretrained UMA-S model naturally inherits the level of theory of the OC20 training set, which is labeled with RPBE functional without spin polarization and without a Hubbard $U$ correction~\cite{chanussot_open_2021}. Therefore, evaluating the pretrained model against the PBE+D3+$U$+spin labels used for fine-tuning would lead to conflating two different level of theory, though the inclusion of Hubbard $U$ correction and spin polarization is expected to better represent the system we are interested in~\cite{behler2005dissociation,behler2008nonadiabatic,wang2011electronic,bondarenko2015polaron,gerosa2016anisotropic}. To evaluate the model's performance under the DFT setting that it was actually trained for, we benchmark it against a DFT reference computed under the OC20's own DFT setting.

For this benchmark, we compute the energies and forces of the R4 test set configurations with DFT in the OC20 setting. We use VASP~\cite{kresse1996efficient,kresse1999ultrasoft} with PAW potentials~\cite{kresse1999ultrasoft} (PBE dataset, version 54) at the RPBE level, non-spin-polarized, and without a Hubbard $U$, matching the protocol used to generate the OC20 dataset~\cite{chanussot_open_2021}. These single-point calculations are managed with the Quantum Accelerator (\texttt{quacc}) workflow library~\cite{rosen2023quacc} so that the input settings reproduce those of OC20. We use the OC20 DFT setting with VASP only when evaluating pretrained model with R4 test set. All the other DFT calculations, including the energy, force, and stress labels used for fine-tuning, is performed in Quantum ESPRESSO at the PBE+D3+$U$+spin level described in Sec.~\ref{sec:methods:ft} (step~3).

\begin{figure}
    \centering
    \includegraphics[width=\linewidth]{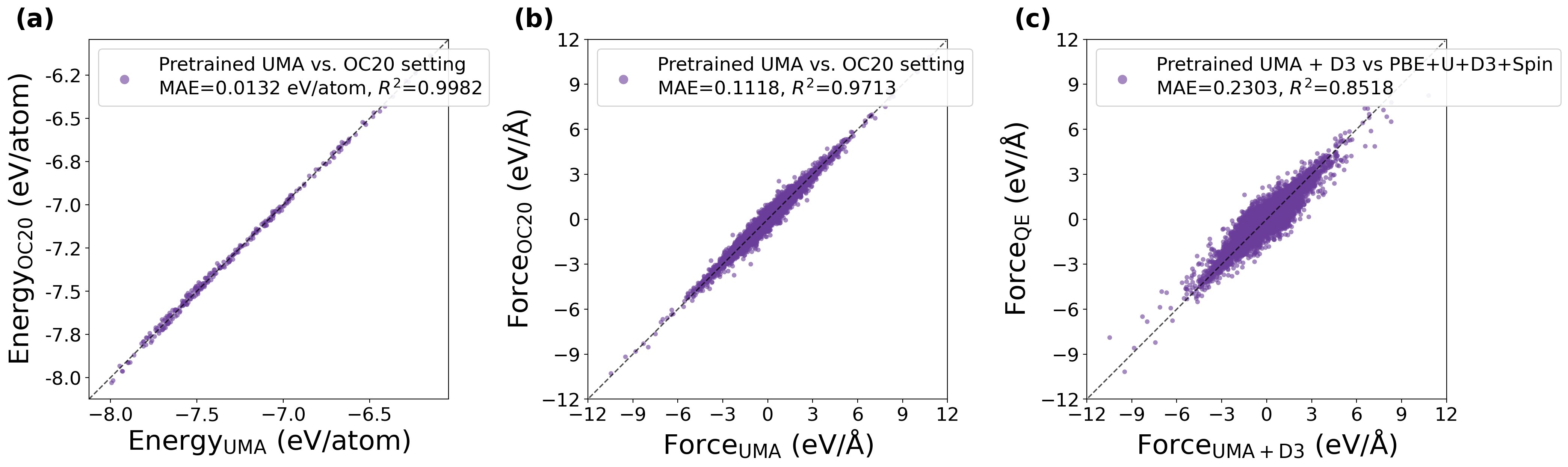}
    \caption{Parity plot of the pretrained UMA-S model evaluated against DFT reference values on the R4 test set. Under its native OC20 task setting (RPBE, non-spin-polarized, no Hubbard $U$), (a)~the energy MAE and (b)~the force MAE are evaluated to be 0.013~eV/atom ($R^{2}{=}0.998$) and 0.112~eV/\AA{} ($R^{2}{=}0.971$), respectively. (c)~The force MAE of the pretrained UMA-S\,+\,D3 model against the PBE+D3+$U$+spin reference is 0.230~eV/\AA{} ($R^{2}{=}0.852$). Energies are total energy per atom.}
    \label{fig:pretrained_parity_oc20}
\end{figure}

Fig.~\ref{fig:pretrained_parity_oc20} shows the resulting parity plots evaluated on the R4 test set. Evaluated against the OC20 setting, the pretrained UMA-S model reaches an energy MAE of $0.013$~eV/atom ($R^{2}{=}0.998$) and force MAE of $0.112$~eV/\AA{} ($R^{2}{=}0.971$). When the same pretrained model, now with D3 dispersion added, is instead evaluated against the PBE+D3+$U$+spin reference, the force MAE rises to $0.230$~eV/\AA{} ($R^{2}{=}0.852$). This inflation reflects the change in functional, dispersion, spin treatment, and Hubbard $U$ between the two references rather than a deficiency of the model.

\subsection{Router layer freezing for LoRA fine-tuning}
\label{app:router-freeze}
In Sec.~\ref{sec:methods:ft}, it was noted that for the LoRA fine-tuning strategy, the MoLE router layer is held frozen for the first 10 epochs, while the full and layer-frozen strategies leave the router trainable throughout. The MoLE strategy used in UMA works by assigning a weight to each expert, which is computed by the router based on the composition of the configuration and the task (OC20, OMat, etc.)~\cite{wood_uma_2026}. During the fine-tuning, the router is also subject to training, which means that the expert weights can be adjusted and therefore be biased toward whichever experts happen to lower the loss on the small fine-tuning set, before the rest of the block has learned to produce the correct outputs for the new W--S--O chemistry. As a result, freezing or unfreezing the router for the first few epochs can yield better fine-tuning performance depending on the fine-tuning strategy.

\begin{figure}
    \centering
    \includegraphics[width=\linewidth]{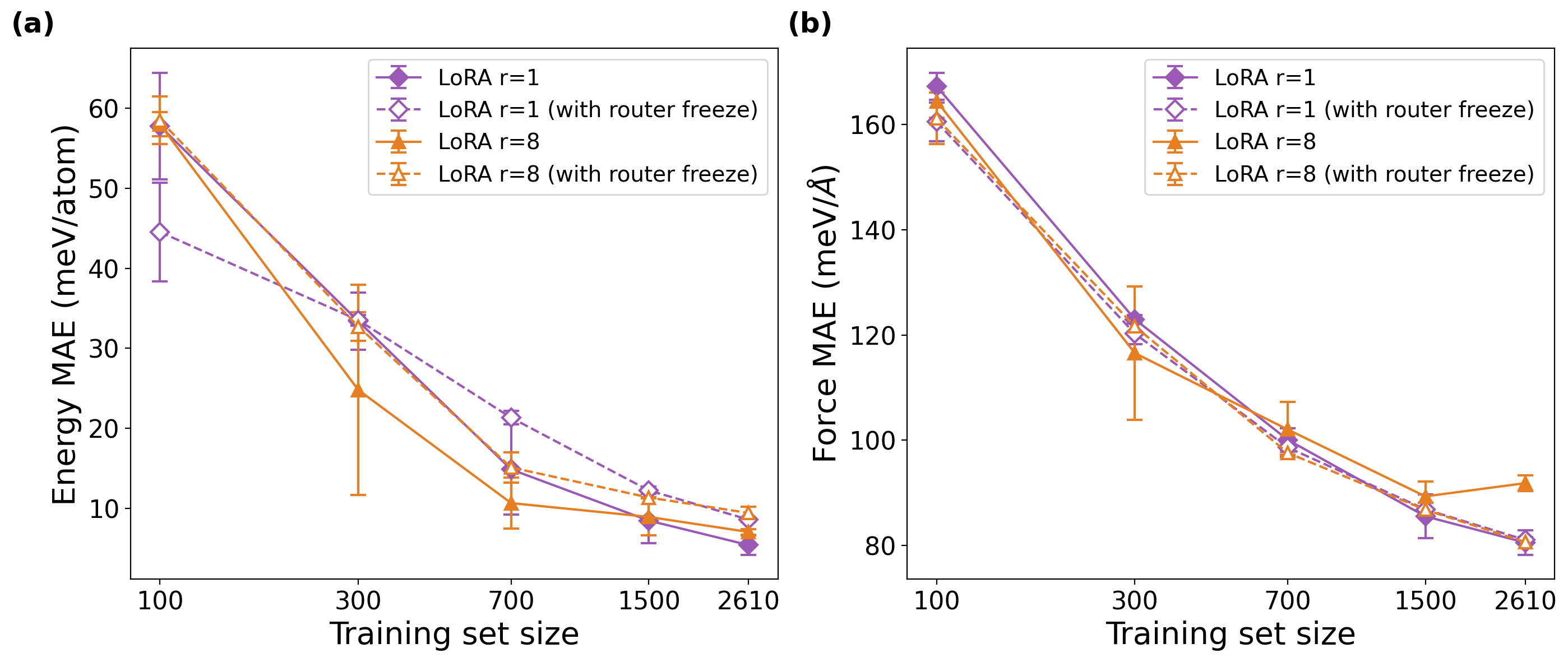}
    \caption{Learning curves showing the effect of holding the MoLE router frozen for the first 10 epochs on LoRA fine-tuning of UMA-S. (a)~Energy MAE and (b)~force MAE on the R4 test set are evaluated as a function of training-subset size for LoRA $r{=}1$ (2.5~M trainable parameters) and LoRA $r{=}8$ (6.8~M trainable parameters), each with and without the 10-epoch router freeze. Solid lines and filled markers are without router freeze; dashed lines and open markers are with router freeze. The error bars are used to indicate the spread of MAEs across the three different resamples. Energies are total energy per atom.}
    \label{fig:learning_curve_router_freeze_lora}
\end{figure}

\begin{figure}
    \centering
    \includegraphics[width=\linewidth]{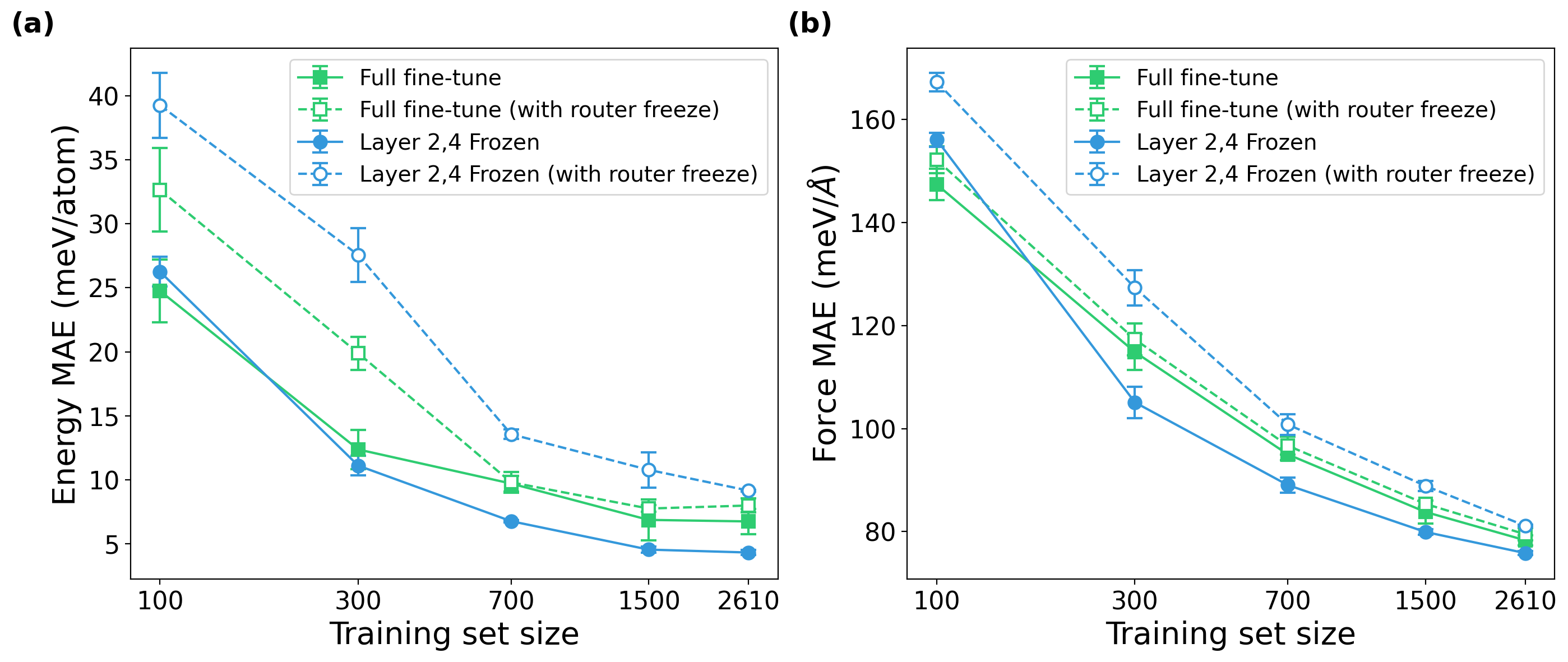}
    \caption{Learning curves showing the effect of holding the MoLE router frozen for the first 10 epochs on the different models with different fine-tuning strategies. Similar to Fig.~\ref{fig:learning_curve_router_freeze_lora}, (a)~Energy MAE and (b)~force MAE on the R4 test set are evaluated as a function of training-subset size for full fine-tuning (146.6~M trainable parameters) and Layer~2,4 freezing (73.4~M trainable parameters), each with and without the 10-epoch router freeze. Solid lines and filled markers are without router freeze; dashed lines and open markers are with router freeze. The error bars are used to indicate the spread of MAEs across the three different resamples. Energies are total energy per atom.}
    \label{fig:learning_curve_router_freeze_full}
\end{figure}

The learning curves of Fig.~\ref{fig:learning_curve_router_freeze_lora} and Fig.~\ref{fig:learning_curve_router_freeze_full} compare the fine-tuning performance with and without the router frozen for the first 10 epochs, across different training-subset sizes and fine-tuning strategies. The difference is most prominent at 100 training configurations, where the LoRA $r{=}1$ energy MAE drops from roughly 60~meV/atom without the router freeze to roughly 45~meV/atom with it (Fig.~\ref{fig:learning_curve_router_freeze_lora}, panel (a)). More importantly, for LoRA $r{=}8$ without the router freeze, increasing the training set size from 1500 to 2610 configurations actually leads to a slight increase in force MAE, from roughly 92~meV/\AA{} to 96~meV/\AA{} (an increase well within the error spread), whereas the router-frozen LoRA $r{=}8$ does not show this. This suggests that router freezing can be important for LoRA fine-tuning. For full fine-tuning and Layer~2,4 freezing, the router-frozen variant has consistently higher energy MAE and force MAE than the non-frozen variant at every training-set size. We therefore apply the 10-epoch router freeze to the LoRA strategies only.

\subsection{Performance of the fine-tuned model on larger simulation cells}
\label{app:OOD_performance}
Because the training configurations, as described in Sec.~\ref{sec:methods:ft}, are from smaller cells than the $6\!\times\!6\!\times\!3$ slab used in the production-scale simulations of Sec.~\ref{sec:methods:sim}, there may be a gap between the training data and the large-cell configurations the model encounters during the production run. Specifically, this difference can arise because the production cells have two fully mobile layers, whereas the training cells have only one. However, directly labeling the configurations from the $6\!\times\!6\!\times\!3$-cell simulation is computationally prohibitive. For that reason, we construct a cell that is larger than the training cells but still tractable to label: a $4\!\times\!4\!\times\!3$ slab, which is larger in both width and height. We take configurations generated from a simulation run with the R3 fine-tuned model and, as in Sec.~\ref{sec:methods:ft}, strip the bottom fixed atoms from the saved configurations and sample them using FPS in the same 50-component SOAP+PCA latent space before labeling. In total, 18 configurations were labeled with the same PBE+D3+$U$+spin DFT protocol and used as a held-out test set, which we refer to as the Round 4b dataset.
 
\begin{figure}
    \centering
    \includegraphics[width=\linewidth]{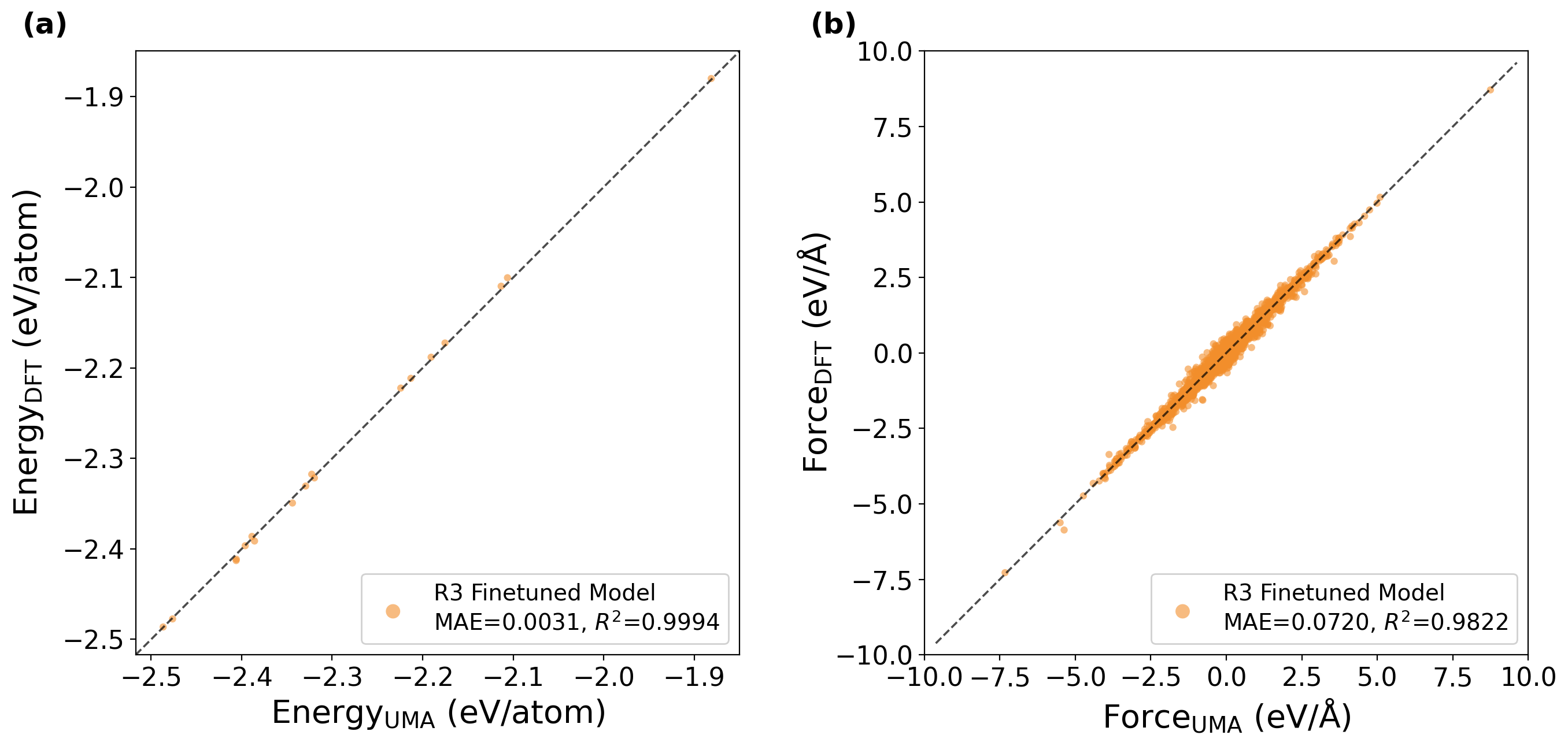}
    \caption{Parity plot of the R3 fine-tuned model against the DFT reference (PBE+D3+$U$+spin) on round 4b dataset, drawn from 15~eV O$^+$/O$_2^+$ bombardment of a larger $4\!\times\!4\!\times\!3$ slab unlike round 4 dataset (described in Sec.~\ref{sec:methods:ft}~(step~1)). The configurations are selected to be maximally diverse from the R1--R3 training set in the configuration-averaged SOAP+PCA space: (a)~energy (MAE${=}3.1\times 10^{-3}$~eV/atom, $R^{2}{=}0.999$) and (b)~force (MAE${=}0.072$~eV/\AA, $R^{2}{=}0.982$). The closer the points are to the dashed line (y$=$x), the better the agreement between the fine-tuned model and DFT. Energies are total energy per atom.}
    \label{fig:round4b_parity}
\end{figure}

\begin{figure}
    \centering
    \includegraphics[width=\linewidth]{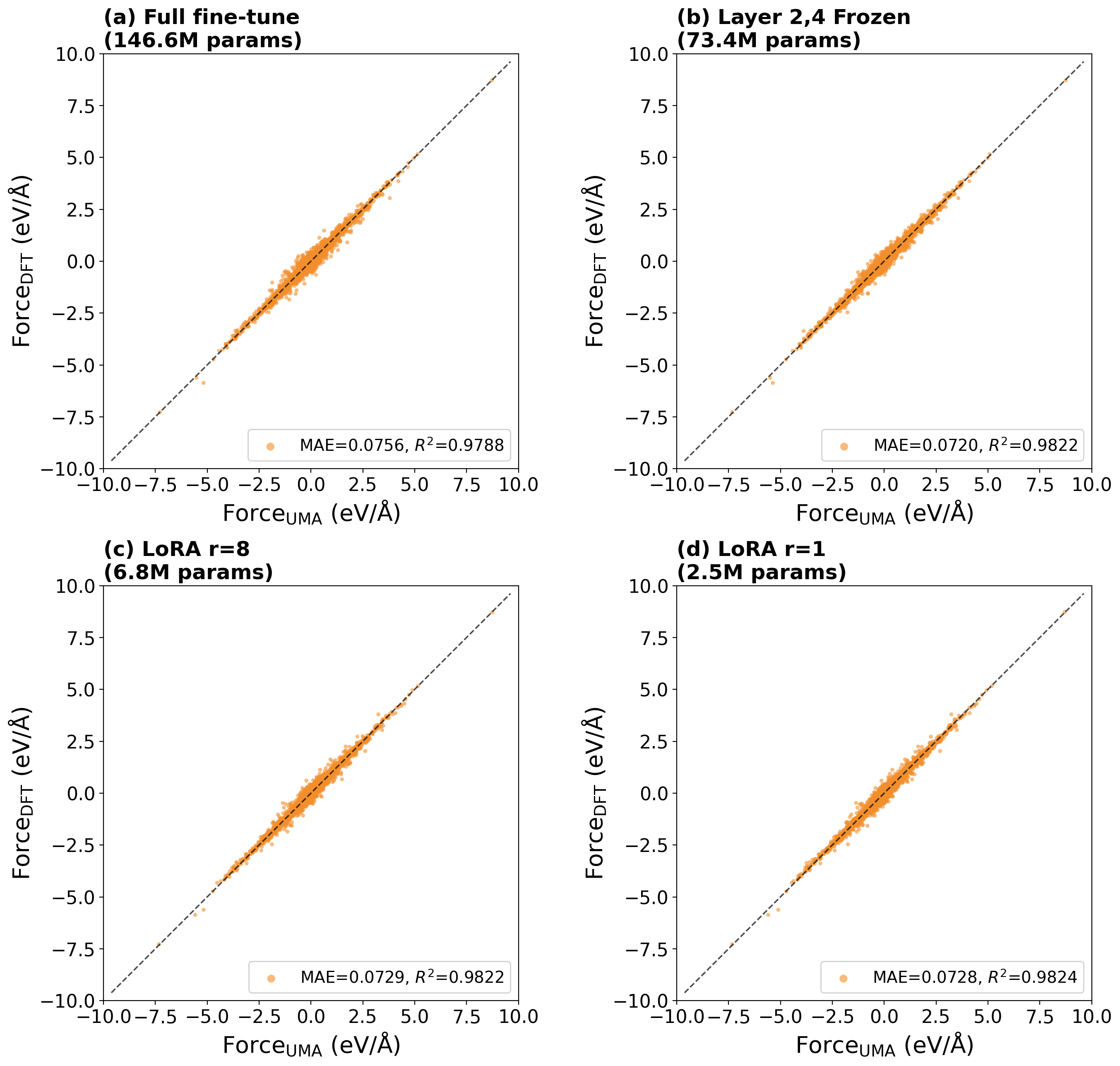}
    \caption{Force parity on the 18-configuration round 4b dataset for the four fine-tuning strategies: (a)~full fine-tuning (146.6~M trainable parameters, MAE${=}0.076$~eV/\AA, $R^{2}{=}0.979$), (b)~Layer~2,4 frozen (73.4~M, MAE${=}0.072$~eV/\AA, $R^{2}{=}0.982$), (c)~LoRA $r{=}8$ (6.8~M, MAE${=}0.073$~eV/\AA, $R^{2}{=}0.982$), and (d)~LoRA $r{=}1$ (2.5~M, MAE${=}0.073$~eV/\AA, $R^{2}{=}0.982$). The round 4b dataset is drawn from 15~eV O$^+$/O$_2^+$ bombardment of a larger $4\!\times\!4\!\times\!3$ slab, unlike round 4 dataset (described in Sec.~\ref{sec:methods:ft}~(step~1)). All four strategies are fine-tuned on the full 2610-configuration R1--R3 training set. The dashed line (y$=$x) marks perfect agreement.}
    \label{fig:round4b_gap_force_parity_methods}
\end{figure}

Fig.~\ref{fig:round4b_parity} shows essentially no difference in the performance of the R3 fine-tuned model between the smaller-cell Round 4 dataset and the larger-cell Round 4b dataset. The energy MAE is $3.1\times 10^{-3}$~eV/atom ($R^{2}{=}0.999$) and the force MAE is $0.072$~eV/\AA{} ($R^{2}{=}0.982$), compared with $4.5\times 10^{-3}$~eV/atom and $0.076$~eV/\AA{} on the R4 test set (Fig.~\ref{fig:r3_finetuned_parity}). This agreement is not specific to the fine-tuning strategy either, as shown in Fig.~\ref{fig:round4b_gap_force_parity_methods}. The resulting force MAEs ($0.076$, $0.072$, $0.073$, and $0.073$~eV/\AA{} for full fine-tuning, Layer~2,4 frozen, LoRA $r{=}8$, and LoRA $r{=}1$, respectively) are within a few percent of one another and of their in-distribution values in Fig.~\ref{fig:birch_finetuning_methods} ($0.079$, $0.076$, $0.080$, and $0.082$~eV/\AA{} for full fine-tuning, Layer~2,4 frozen, LoRA $r{=}8$, and LoRA $r{=}1$, respectively). This indicates that the model transfers to larger cells without loss of accuracy, so the smaller cells used for the training and test sets are sufficient.

\subsection{Fine-tuning performance over different fine-tuning methods}
\label{app:results-finetune_methods}
The round-over-round results in Sec.~\ref{sec:results-finetune_rounds} use a single fine-tuning strategy, namely layer freezing, with the layer to freeze chosen by hyperparameter tuning. We now investigate how much the converged accuracy depends on that choice. This is critical because fine-tuning operates in a much smaller-data regime than pretraining: UMA-S was pretrained on $500$~million labeled atomic configurations spanning broad chemistry and structure space~\cite{wood_uma_2026}, while our cumulative R1--R3 fine-tuning set contains $2610$ configurations, more than five orders of magnitude smaller. This shift in data-to-parameter ratio is the standard motivation for considering parameter-efficient fine-tuning strategies alongside full fine-tuning, since lighter strategies can in principle preserve more of the broadly-transferable representations that the pretrained model already encodes while still adapting the model to the target chemistry. The four strategies considered here span nearly two orders of magnitude in trainable-parameter count: full fine-tuning at $146.6$~M, layer freezing at $73.4$~M, LoRA $r{=}8$ at $6.8$~M, and LoRA $r{=}1$ at $2.5$~M. LoRA provides a particularly informative comparison point because constraining the fine-tuning update to a rank-$r$ correction on the expert weights limits the expressiveness of the update and acts as implicit regularization, with the effect becoming stronger as $r$ is reduced~\cite{hu_lora_2021}. Comparing the four strategies therefore lets us ask how the converged accuracy depends on trainable-parameter count and on the strength of implicit regularization, and whether the LoRA variants reach the accuracy of full or layer-frozen fine-tuning despite their much smaller parameter count.
 
\begin{figure}
    \centering
    \includegraphics[width=\linewidth]{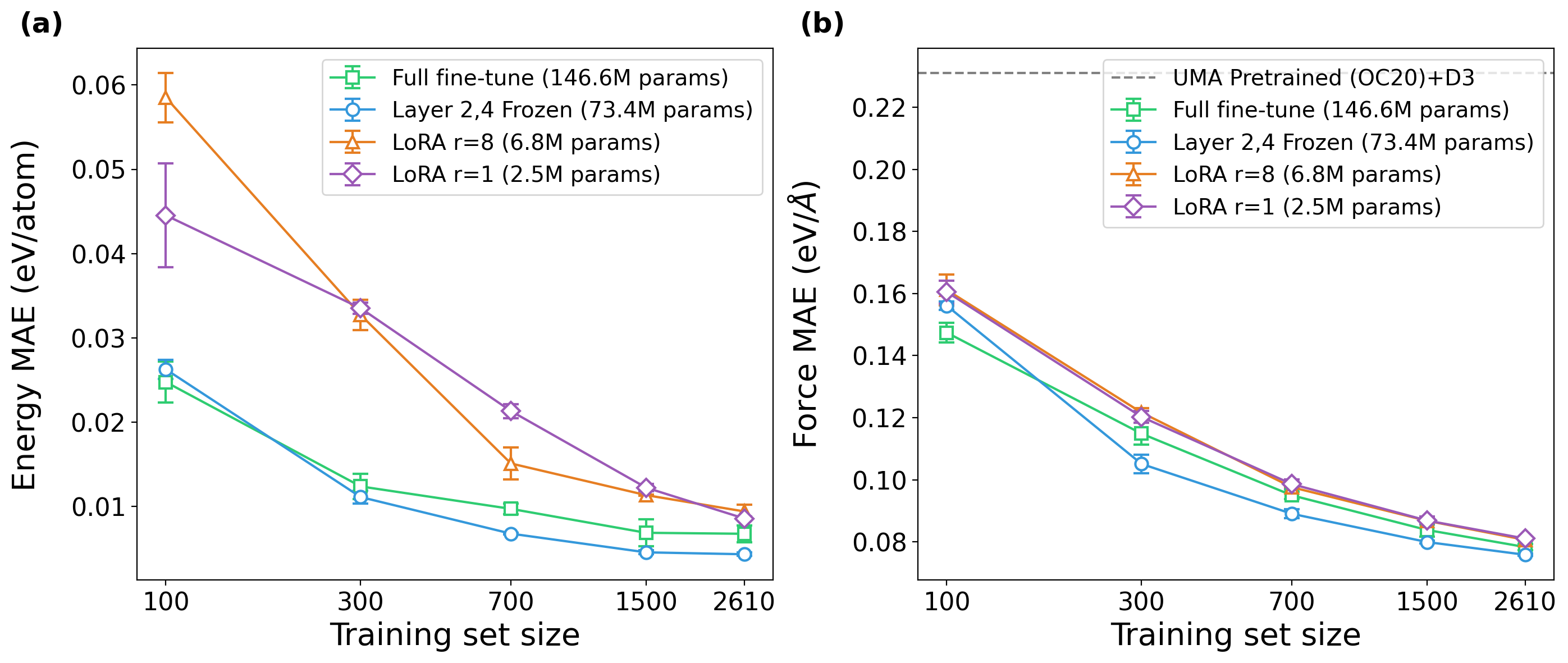}
    \caption{Learning curves for four different fine-tuning strategies are displayed: full fine-tuning, Layer~2,4 freezing, LoRA $r{=}8$, and LoRA $r{=}1$. The (a)~Energy MAE and (b)~force MAE on the R4 test set as a function of training-subset size of 100, 300, 700, 1500, 2610 configurations across R1 to R3 dataset are evaluated. For each training-subset size, the training subset is resampled three times, and the error bars are used to display the spread of the MAE results from those three resamples. The horizontal line in (b) marks the force MAE of the pretrained UMA-S~(OC20)+D3 model. The number of trainable parameters for each fine-tuning strategy is displayed in the legend. Energies are total energy per atom.}
    \label{fig:learning_curve}
\end{figure}

Fig.~\ref{fig:learning_curve} shows the energy MAE and force MAE on the R4 test set as a function of the size of the training subset for each of the four strategies. Two observations emerge. First, at every training-set size, all four strategies improve on the pretrained UMA-S~(OC20)+D3 baseline (horizontal dashed line in panel (b)) by more than a factor of two, including LoRA $r{=}1$, which has only $\sim 1.7\%$ of the trainable parameters of full fine-tuning. Second, the four strategies converge to nearly indistinguishable force MAE at the full $2610$-configuration training-set size, with the layer-freezing strategy reaching the lowest error and the LoRA variants closing most of the gap. The convergence of the four curves at the right edge of the plot is the operative result for the present study: at the training-set size we actually use for the production model, the strategy choice changes the converged accuracy by only a few percent.

\begin{figure}
    \centering
    \includegraphics[width=\linewidth]{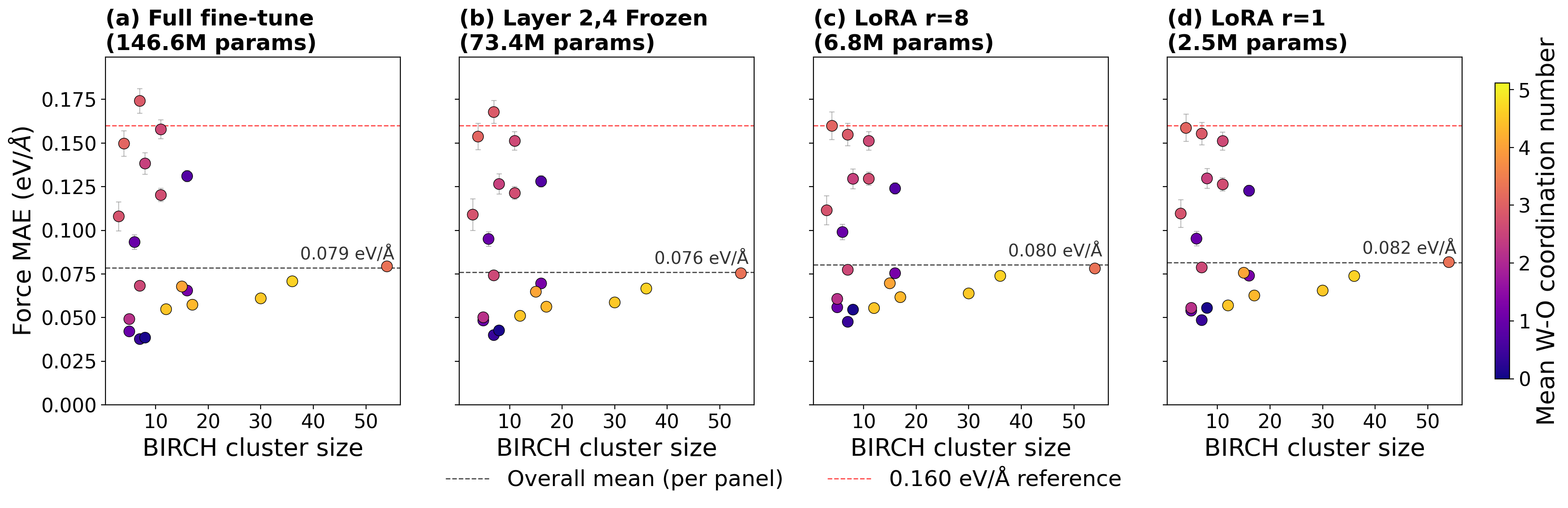}
    \caption{Local-environment-resolved force MAE on the R4 test set across different fine-tuning strategies are displayed: (a)~full fine-tuning, (b)~Layer~2,4 frozen, (c)~LoRA $r{=}8$, and (d)~LoRA $r{=}1$. All four strategies are fine-tuned on all 2610 training set configurations from R1--R3 dataset. The test set configurations are grouped into 20 BIRCH clusters from the SOAP+PCA space defined in Sec.~\ref{sec:methods:ft}~(step~2). Each marker is for each cluster, and they are plotted based on the number of configurations within the cluster and mean force MAE of the cluster, with the spread of force MAE across the cluster displayed with error bars. They are also colored based on their mean W-O coordination number of the cluster's atoms. The black dashed line is the per-panel mean force MAE (0.079, 0.076, 0.080, and 0.082~eV/\AA{} in panels a--d), and the red dashed line at 0.160~eV/\AA{} is a diagnostic threshold for per-cluster force MAE. The trainable parameters are indicated in the parentheses for each fine-tuned model.}
    \label{fig:birch_finetuning_methods}
\end{figure}

To see where the residual force error sits within each strategy, we again apply the BIRCH-clustered analysis, this time with the panels splitting by fine-tuning strategy rather than by round (Fig.~\ref{fig:birch_finetuning_methods}). The per-panel means ($0.079$, $0.076$, $0.080$, $0.082$~eV/\AA{} for full fine-tuning, Layer~2,4 frozen, LoRA $r{=}8$, and LoRA $r{=}1$, respectively) are the force MAE from a single training-subset resample at the full $2610$-configuration size, i.e., one of the replicates that contributes to the rightmost error bar of the learning curves in Fig.~\ref{fig:learning_curve} rather than the resample-averaged value shown there. They are reported here only to label each panel; the new information in the figure is the within-panel distribution of those errors over local environments. The marker patterns within each panel are similar: in every strategy, mid-W-O-coordination clusters (mean W-O of roughly 2--4) carry the largest residual force error, and a small number of these sit above the $0.160$~eV/\AA{} diagnostic line.

Interestingly, although LoRA $r{=}1$ has the highest mean force MAE of the four strategies, its highest-error clusters are modestly lower than the corresponding highest-error clusters of the other three strategies. This suggests that the LoRA $r{=}1$ parameterization, which restricts the fine-tuning update to a rank-1 correction on each expert (Sec.~\ref{sec:methods:ft}), may be slightly less prone to overfitting to the bulk of the training distribution at the expense of the tails. The difference is not dramatic, but it does indicate that for the W--S--O chemistry studied here the additional capacity of full or layer-frozen fine-tuning is not what is limiting the converged accuracy. The dominant bottleneck is instead the coverage of the mid-coordination, partially-oxidized W environments in the training set, exactly as identified in Sec.~\ref{sec:results-finetune_rounds}. Since the residual error is dominated by under-represented local environments rather than by limited model capacity, further accuracy gains on this system are more likely to come from additional rounds of MD-driven configuration sampling than from increasing the number of trainable parameters at fine-tuning time.

\subsection{Etched sulfur species}
\label{app:etched_S}
\begin{figure}
    \centering
    \includegraphics[width=\linewidth]{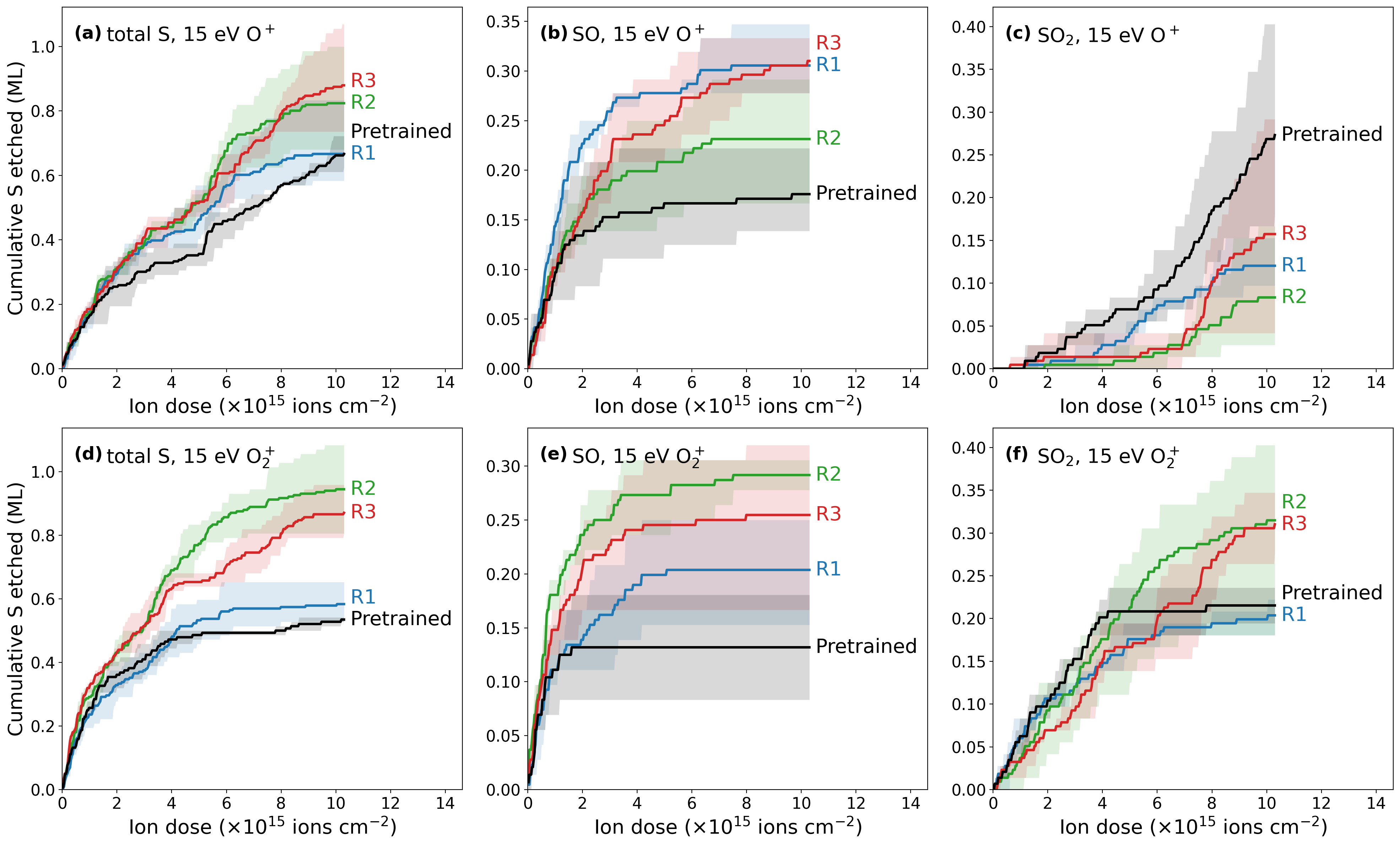}
    \caption{Cumulative sulfur etched as a function of ion dose for the pretrained UMA-S model and the R1, R2, and R3 fine-tuned models. The top row, panels (a)-(c), shows results for 15 eV O$^+$ bombardment, while the bottom row, panels (d)-(f), shows the result for 15 eV O$_2^+$ bombardment. Panels (a) and (d) show total etched S; panels (b) and (e) show S etched as SO; and panels (c) and (f) show S etched as SO$_2$, the two most prominent S-containing etch products. The etched S content is reported in terms of monolayers (ML), which is defined as $1\,\mathrm{ML} = 72$~S atoms, corresponding to two S atoms per W in WS$_2$. The solid line represents the average trend over the three different replicates, and the shaded band shows the minimum and maximum across the three replicates.}
    \label{fig:etched_S}
\end{figure}

Chemisorbed S and O coverage are the observables of interest, but they may plausibly be insensitive to model quality. We therefore tracked the sulfur-containing etched products during the production-scale runs of Sec.~\ref{sec:results-md}. A molecule counts as etched when the cluster removal method of Sec.~\ref{sec:methods:sim} picks it up as a small, W-free fragment moving away from the slab. Physisorbed species that remain on the surface are therefore excluded. Fig.~\ref{fig:etched_S} shows the total etched S and the two most abundant species, SO and SO$_2$. The remaining etched products are distributed across S, S$_2$, S$_2$O, and SO$_3$. The R2 and R3 fine-tuned models track each other closely in most panels, with SO under O$^+$ bombardment (panel (b)) is the clearest exception. The pretrained model etches less total sulfur than either, so more of the non-chemisorbed sulfur stays physisorbed. R2 and R3 differ far less than R2 differs from R1 and R1 from the pretrained model. Further rounds are unlikely to change this observable much.
 
\end{document}